\documentclass[twocolumn,aps,floatfix,superscriptaddress,pra,nobibnotes,longbibliography]{revtex4-1}

\usepackage{graphicx}
\usepackage{amsmath,amsthm,amssymb,amsfonts}

\usepackage[USenglish]{babel}
\usepackage[bbgreekl]{mathbbol}
\usepackage{color}
\usepackage[utf8]{inputenc}

\usepackage[colorlinks=true,linkcolor=blue,citecolor=blue]{hyperref}
\usepackage{accents}
\newlength{\dhatheight}

\definecolor{cg}{rgb}{0.0, 0.5, 0.0}

\def\d{\mathrm{d}}
\def\rmi{\mathrm{i}}

\begin{document}

\title{Supersolid phases of Rydberg-excited bosons on a triangular lattice}

\author{Jaromir Panas}
\email{panas@th.physik.uni-frankfurt.de}
\affiliation{Institut f\"ur Theoretische Physik, Goethe-Universit\"at, 60438 Frankfurt am Main, Germany}

\author{Mathieu Barbier}
\affiliation{Institut f\"ur Theoretische Physik, Goethe-Universit\"at, 60438 Frankfurt am Main, Germany}

\author{Andreas Gei{\ss}ler}
\affiliation{Institut f\"ur Theoretische Physik, Goethe-Universit\"at, 60438 Frankfurt am Main, Germany}
\affiliation{ISIS, University of Strasbourg and CNRS, 67000 Strasbourg, France}

\author{Walter Hofstetter}
\affiliation{Institut f\"ur Theoretische Physik, Goethe-Universit\"at, 60438 Frankfurt am Main, Germany}

\date{\today}

\begin{abstract}
Recent experiments with ultracold Rydberg-excited atoms have shown that long-range interactions can give rise to spatially ordered structures. Observation of crystalline phases in a system with Rydberg atoms loaded into an optical lattice seems also within reach. Here we investigate a bosonic model on a triangular lattice suitable for description of such experiments. Numerical simulations based on bosonic dynamical mean-field theory reveal a rich phase diagram with different supersolid phases. Comparison with the results obtained for a square lattice geometry shows qualitatively similar results in a wide range of parameters, however, on a triangular lattice we do not observe the checkerboard supersolid. Moreover, unlike on a square lattice we did not find a phase transition from uniform superfluid to supersolid induced by increase of the hopping amplitude on a triangular lattice. Based on our results we propose an intuitive interpretation of the nature of different supersolid phases. We also propose parameters for the experimental realization.
\end{abstract}


\maketitle

\section{Introduction}
\label{sec:intro}

A supersolid is a phase with simultaneously broken $U(1)$ and translational symmetry of the system. Since the first time it was theoretically discussed~\cite{onsager1956,andreev1969,chester1970,leggett1970} it has proven difficult to realize in experiment. So far it was only observed for ultracold bosons in optical cavities, where the light mode mediates long-range interaction~\cite{leonard2017,landig2016,klinder2015}. Other paths to obtain supersolids are intensively studied. One promising experimental approach involves dipolar quantum gases loaded into an optical lattice~\cite{lahaye2009}. The advantage of this approach is that the resulting system is highly tunable and accurately described by the extended~\cite{batrouni2000,goral2002} version of the Bose-Hubbard model~\cite{gersch1963,fisher1989}.

One of the first studies of the extended Bose-Hubbard model in the context of supersolids involved hard-core bosons on a triangular, frustrated lattice with nearest-neighbor interactions~\cite{wessel2005,heidarian2005,melko2005,boninsegni2005}. In these numerical quantum Monte Carlo (QMC) studies several different phases were observed, including a superfluid, an insulating density-wave, and two supersolid phases. Interestingly, the frustration of the lattice was found to be essential for the formation of the supersolid and for preventing phase separation, which is observed on the square lattice~\cite{batrouni2000}.

To obtain a supersolid phase on a square lattice with a hard-core constraint one needs long-range interaction~\cite{otterlo1995,ohgoe2011}. Studies within mean-field~\cite{goral2002,yi2007} and QMC~\cite{capogrosso2010} have shown that supersolid phases exist in square and cubic lattice models with dipolar interactions between atoms. Long-range interaction also leads to other interesting phenomena, such as the appearance of multiple insulating density-wave phases with commensurate filling in the limit of small hopping amplitude~\cite{capogrosso2010,schachenmayer2010}. The resulting phase diagram has features similar as the devil's staircase in the Ising model~\cite{bak1982,lauer2012,rademaker2013}.

Experimentally, long-range interactions of the van der Waals type can be realized, e.g. by exciting atoms to high principal quantum number Rydberg states~\cite{gallagher1988,singer2005}. The advantage of this approach is that one can tune the strength of the long-range interaction through an appropriate choice of the Rydberg state. Rydberg atoms also have applications in other fields such as quantum information~\cite{saffman2010}, understanding quantum critical behavior~\cite{weimer2008}, molecule~\cite{bendkowsky2009,niederprum2015} and superatom~\cite{weber2015,zeiher2015} formation, among many other~\cite{browaeys2016}. Moreover, loading of Rydberg atoms into an optical lattice and the observation of emerging ordered structures have been achieved experimentally~\cite{schauss2012,schauss2015,zeiher2016,zeiher2017,schauss2018}, although so far only in the frozen-limit of a deep lattice potential. In contrast, theoretical studies of the corresponding models have been performed both in the frozen-limit and for itinerant atoms. Main results include the study of self-organization of Rydberg excitations in a lattice~\cite{rademaker2013,vermersch2015}, phase diagrams and effects of finite hopping amplitude~\cite{lauer2012,saha2014,geissler2017,li2018}, spectral properties of different phases~\cite{geissler2018}, and effects of dissipation~\cite{ray2016,barbier2018}. However, these studies focused on the square lattice geometry without considering effects of frustration.

In this work we aim at extending the study to the triangular lattice, frustrated with respect to formation of a checkerboard solid, and at understanding the influence of system's geometry on the ground state by comparing our new results to those obtained for the square lattice~\cite{geissler2017}. In Sec.~\ref{sec:sec2} we introduce the model and briefly discuss our variant of the real-space bosonic dynamical mean-field theory (B-DMFT) method used in the simulations. In Sec.~\ref{sec:sec3} we present and discuss the results. Sec.~\ref{sec:sub1sec3} is aimed at understanding effects of the lattice geometry by studying the phase diagrams and properties of observed phases. In Sec.~\ref{sec:sub2sec3} we compare results of B-DMFT and static mean-field theory to estimate the relevance of local quantum fluctuations. In Sec.~\ref{sec:sub3sec3} we propose an experimental scheme for minimizing the destructive influence of dissipation induced by coupling to the environment. In Sec.~\ref{sec:end} our findings are summarized.

\section{System and method}
\label{sec:sec2}
\subsection{Model}\label{sec:sub1sec2}

We choose a model suitable to describe experiments with bosonic alkali atoms, e.g. $^{87}$Rb, loaded into a triangular optical lattice~\cite{becker2010}, and coupled (by an additional laser field) to a Rydberg state with high principal quantum number~\cite{schauss2012,zeiher2016}. For each lattice site $i$ we introduce bosonic annihilation operators $\hat{a}^{\phantom\dag}_i$ of an atom in its ground state and $\hat{b}^{\phantom\dag}_i$ of an atom in its highly excited Rydberg state. The corresponding Hamiltonian reads~\cite{saha2014,geissler2017}
\begin{equation}\label{eq:fullH}
\hat{H} = \hat{H}_{kin} +\hat{H}_{vdW} + \sum_{i}\left(\hat{H}_{loc,i} + \hat{H}_{R,i} \right).
\end{equation}
The summation runs over the $N$ lattice sites of the system. In the end we take the thermodynamic limit of $N\to \infty$ assuming the system to be composed of periodically recurring unit cells of finite size $N_{uc}$.

$\hat{H}_{kin}$ represents the kinetic energy of atoms tunneling between neighboring lattice sites
\begin{equation}\label{eq:kinH}
\hat{H}_{kin} = -J\sum_{\langle i,j\rangle }\left(\hat{a}^\dag_{i}\hat{a}^{\phantom\dag}_{j} + \eta \hat{b}^\dag_{i}\hat{b}^{\phantom\dag}_{j} \right).
\end{equation}
Here $J$ is the hopping amplitude, and $\eta$ represents the ratio between the hopping amplitude of excited state atoms to that of ground state atoms. $\langle i,j\rangle$ indicates summation over nearest neighbors $i$ and $j$. Is is useful to introduce the connectivity $z$ of the lattice, which is the number of nearest-neighbors for any site. For the triangular lattice $z=6$.

$\hat{H}_{vdW}$ represents the van der Waals interaction between two excited state atoms and is given by
\begin{equation}\label{eq:vdwH}
\hat{H}_{vdW} = \frac{V_{vdW}}{2}\sum_{i\neq j} \frac{\hat{n}_{e,i} \hat{n}_{e,j}}{|\mathbf{i}-\mathbf{j}|^6},
\end{equation}
where $|\mathbf{i}-\mathbf{j}|$ is the Euclidean distance between lattice sites $i$ and $j$ divided by lattice spacing $a$, $\hat{n}_{e,i} = \hat{b}^\dag_i \hat{b}^{\phantom\dag}_i$ is the number operator at site $i$ for the excited bosons, $V_{vdW}$ is the van der Waals interaction strength, which is given by $V_{vdW}=C_6/a^6$ with $C_6$ being van der Waals coefficient~\cite{singer2005}.

$\hat{H}_{loc,i}$ is a local part of the Hamiltonian (for site $i$) describing the chemical potential and the onsite interaction. It is given by
\begin{equation}\label{eq:locH}
\begin{split}
\hat{H}_{loc,i} & = \frac{U}{2} \left( \hat{a}^\dag_i \hat{a}^\dag_i \hat{a}^{\phantom\dag}_i \hat{a}^{\phantom\dag}_i 
+ 2 \lambda \hat{a}^\dag_i\hat{b}^\dag_i\hat{b}^{\phantom\dag}_i\hat{a}^{\phantom\dag}_i 
+ \tilde{\lambda} \hat{b}^\dag_i\hat{b}^\dag_i\hat{b}^{\phantom\dag}_i\hat{b}^{\phantom\dag}_i \right) \\
& - \mu\left( \hat{n}_{g,i} + \hat{n}_{e,i}\right),
\end{split}
\end{equation}
with $\hat{n}_{g,i} = \hat{a}_i^\dag\hat{a}^{\phantom\dag}_i$. The parameters $U$, $\lambda U$ and $\tilde{\lambda} U$ describe the local interaction strength between two ground state atoms, ground state atom and excited state atom, and two excited state atoms, respectively. $\mu$ is the chemical potential of an external thermal reservoir, since we work in the grand canonical ensemble.

The last term in the Hamiltonian, the Rabi term $\hat{H}_{R,i}$, describes coupling between ground and excited state atoms, induced by the driving with an additional laser field. Within the rotating wave approximation (RWA) this contribution to the Hamiltonian is given by
\begin{equation}\label{eq:rabH}
\hat{H}_{R,i} = \frac{\Omega}{2} \left( \hat{b}^\dag_i \hat{a}_i + \hat{a}^\dag_i\hat{b}_i\right) - \Delta \hat{n}_{e,i}.
\end{equation}
Here $\Omega$ is the Rabi frequency, and $\Delta$ the detuning of laser frequency from that of the atomic transition which we consider.

In the following we set $\hbar=k_B=1$ and use the Rabi frequency $\Omega$ as the unit of energy, unless stated otherwise. We assume the system is in thermal equilibrium at zero temperature.

In our model we set $\lambda,\tilde{\lambda}\gg 1$ leading to a hard-core constraint for excited state atoms~\cite{geissler2017}. Rydberg atoms are susceptible to formation of molecules~\cite{manthey2015}, which are not trapped by the lattice potential and therefore lead to a high two-body loss rate. This in turn leads to a hard-core constraint due to the quantum Zeno effect~\cite{misra1977,garcia2009,vidanovic2014}.

We also set the value of $\eta=0$ which translates to immobile Rydberg atoms. The effect of non-vanishing $\eta$ was considered in~\cite{geissler2017} and only small changes in the values of observables were observed. This is not surprising as the excited atoms interact strongly via the van der Waals interaction, which leads either to a very small fraction of excited atoms or to crystalline order where kinetic processes are suppressed.

\subsection{Method}
\label{sec:sub2sec2}

We perform calculations with two methods: (i) a Gutzwiller (static) mean-field approximation, described in detail in~\cite{fisher1989,barbier2018}, and (ii) the bosonic dynamical mean-field theory (B-DMFT)~\cite{byczuk2008}. Both methods are based on self-consistency and on mapping of the lattice problem onto a set of local impurity problems. In order to be able to do the latter we treat the non-local interaction term within the Hartree approximation~\cite{negele1998}
\begin{equation}\label{eq:Hartree}
\begin{split}
\hat{H}_{vdW} & = \frac{V_{vdW}}{2}\sum_{i\neq j} \frac{\hat{n}_{e,i} \hat{n}_{e,j}}{|\mathbf{i}-\mathbf{j}|^6} \\
& \approx V_{vdW}\sum_{i\neq j} \left(\hat{n}_{e,i} - \frac{\langle\hat{n}_{e,i}\rangle}{2}\right) \frac{\langle\hat{n}_{e,j}\rangle}{|\mathbf{i}-\mathbf{j}|^6}.
\end{split}
\end{equation}
Moreover, both methods are implemented within a real-space approach, which allows to study arbitrary periodically recurring ordered structures.

Below we will outline the main steps of the B-DMFT approach, referring the reader to~\cite{geissler2017} for a more detailed discussion. The Gutzwiller mean-field technique may be viewed as a limiting case of B-DMFT and, therefore, it follows similar steps.

\subsubsection{Frozen-limit}

To efficiently perform the B-DMFT calculations we first need to predict what kind of self-organized structures may emerge in the system due to the long-range interaction. We therefore first perform calculations for the frozen gas with $J=0$. In this limit at unit filling one can map the problem onto an effective spin model~\cite{weber2015,zeiher2016}, which however is still not trivial to solve on an infinite two-dimensional triangular lattice. We therefore perform another simplification, assuming a negative value of the chemical potential $\mu<0$, which in the frozen-limit leads to a dilute crystal. 

Owing to the negative chemical potential and zero temperature the bosons can reside in the lattice only when their energy is sufficiently lowered by the Rabi term $\hat{H}_{R}$. When this is the case, the ground state of the system will be a spatially periodic structure with optimal balance between the distribution of bosons in the system and the strength of interaction between them. One can efficiently find free energies of many metastable, spatially periodic states. Each such state is characterized by two spanning vectors $\mathbf{v}_1$ and $\mathbf{v}_2$, see Fig.~\ref{fig:frozen} (top). The bosons in the lattice reside only on the sites related by translations defined be these two vectors, forming a sublattice of the underlying triangular lattice. At each occupied lattice site there is exactly one boson $\langle\hat{n}_g +\hat{n}_e\rangle = 1$ in a superposition of a ground and excited state. Comparing the free energies of these metastable ordered states, one can determine the ground state of the system. Structures associated in the frozen-limit with a ground state for a certain value of the detuning are considered in the B-DMFT calculation later on.

We note that in the frozen-limit a simple expression was found for the critical value of the detuning at which the system undergoes a phase transition to vacuum~\cite{geissler2017}. This expression can be easily extended beyond the frozen-limit as the transition between the vacuum and a very dilute gas, in which van der Waals type interactions are negligible, can be treated as a single-particle problem (see also App.~\ref{app:scaling}). One finds a critical value of the hopping amplitude as a function of the chemical potential, detuning, Rabi frequency and connectivity
\begin{equation}\label{eq:vacuum}
z J_c = \frac{\Omega^2+\Delta^2 -(2\mu+\Delta)^2}{4(\mu+\Delta)}.
\end{equation}

We emphasize that the frozen-limit approach taken here is used primarily to predict most relevant structures for the further B-DMFT and static mean-field calculations. While in the frozen limit we do neglect certain orderings reported for lattice gas models that could not be described with just two Bravais vectors~\cite{lee2002,rademaker2013}, we still can recover some of them within B-DMFT (up to certain wavelengths of the structure) because there each site of the unit-cell is treated independently. We are not able to describe disordered, e.g. glassy, phases.

\subsubsection{B-DMFT}

The results of the frozen-limit allow to select the relevant ordered structures and thus to reduce the number of B-DMFT calculations by selecting only those pairs of spanning vectors $(\mathbf{v}_1,\mathbf{v}_2)$ which correspond to some ground state of the system in the frozen-limit. Each pair $(\mathbf{v}_1,\mathbf{v}_2)$ defines a unit cell with $N_{uc}$ sites that recurs periodically in the system. Within this unit cell a separate quantum impurity problem corresponds to each site. These impurity problems might have different parameters and solutions, resulting in different values of local observables, such as the condensate order parameter for ground $\phi_{i,g}=\langle \hat{a}_i\rangle$ and excited state $\phi_{i,e}=\langle\hat{b}_i\rangle$ bosons, expectation value of the occupation of ground $\langle\hat{n}_{g,i}\rangle$ and excited state $ \langle\hat{n}_{e,i}\rangle$ bosons, connected local Green functions, self-energies, etc. As impurity solver within the B-DMFT calculations we apply the exact diagonalization method~\cite{hubener2009,hu2009,geissler2017}.

Within B-DMFT one needs to define a set of self-consistency equations~\cite{georges1996,byczuk2008}. The first one is given by the local Dyson equation, relating the local interacting connected Green function $\mathbf{G}$, local Weiss field $\mathbf{\mathcal{G}}$ and the self-energy $\mathbf{\Sigma}$. It reads
\begin{equation}\label{eq:locDys}
\mathbf{\mathcal{G}}^{-1}_i (\rmi \omega_n) = \mathbf{G}^{-1}_i (\rmi \omega_n) + \mathbf{\Sigma}_i (\rmi\omega_n).
\end{equation}
Note that each object here is a $4\times 4$ matrix, since there are two components due to the Nambu notation for bosonic Green functions~\cite{byczuk2008} and two components due to two types of bosons (ground and excited state) in the lattice.

The second self-consistency equation involves the condensate order parameter and reads
\begin{equation}\label{eq:condensate_field}
\mathbf{\Psi}_i = \left( \mathbf{g}_i^0 (0) - \mathbf{\mathcal{G}}_i(0) \right) \mathbf{\Phi}_i +  \sum_{j:\langle i,j\rangle }J \mathbf{\Phi}_j,
\end{equation}
where $\mathbf{g}^0_i$ is the Green function of the non-interacting lattice site $i$ decoupled from the rest of the lattice, $\mathbf{\Psi}_i$ is the vector determining the condensate mean-field to which the impurity $i$ is coupled and $\mathbf{\Phi}_j$ is the vector determining the order parameter at site $j$ (which is calculated in the impurity problem). The summation runs over all $j$ which are nearest-neighbors of site $i$.

In standard B-DMFT the last self-consistency equation would be given by the lattice Dyson equation~\cite{byczuk2008}. Here, however, due to the complexity of the problem and the large spatial structures considered we used a simpler one. In our approximate approach we determine the Weiss field according to
\begin{equation}\label{eq:betheSC}
\mathbf{\mathcal{G}}_i (\rmi \omega_n)= \mathbf{g}^0_i (\rmi\omega_n) - \sum_{j:\langle i,j\rangle} J^2 \mathbf{G}_j (\rmi \omega_n).
\end{equation}
Such self-consistency equation would become exact in the limit of the infinite connectivity Bethe tree~\cite{byczuk2008}. In finite spatial dimension it amounts to neglecting: (i) the effect which removing a site from the lattice has on the lattice Green functions, (ii) correlations between different neighbors of the impurity. A similar (though not identical) self-consistency equation has been successfully used for lattices in finite dimensions in the context of real-time dynamics~\cite{strand2015}. In App.~\ref{app:supplementary results} we elaborate on the effect of this approximation.

Having obtained the self-consistent solution one can use the local quantities calculated in the impurity problems to determine values of non-local quantities. Most importantly one can calculate the free energy~\cite{geissler2017}. Note that we have included the chemical potential into the Hamiltonian~\eqref{eq:fullH} and are working at zero temperature, therefore the free energy per lattice site is given by $f=\langle \hat{H} \rangle/N$.

\section{Results}
\label{sec:sec3}

In this section we present the results of our calculations. In Sec.~\ref{sec:sub1sec3} we set the system parameters to be comparable to those used in~\cite{geissler2017}, where the same model on a square lattice has been studied with B-DMFT. This allows us to investigate how the triangular lattice geometry affects the behavior of the system. We also study the nature of different phases observed. In Sec.~\ref{sec:sub2sec3} we compare the static mean-field and the B-DMFT results to estimate the significance of local quantum fluctuations. These two sections are aimed at giving a better understanding of phases emerging due to the competition of long-range interaction and kinetic processes on a triangular lattice. In Sec.~\ref{sec:sub3sec3} we study a system with experimentally more feasible parameters. We investigate the possibility of observing supersolid phases in a triangular optical lattice with Rydberg atoms. To minimize dissipative effects we follow the idea suggested in~\cite{geissler2018} of using an inhomogeneous profile of the Rabi laser.

\subsection{Phase diagram}
\label{sec:sub1sec3}

We choose the following parameters of our system: $V_{vdW}=100\Omega$, $U=0.1\Omega$, $\mu=-0.25\Omega$. As discussed earlier we also set $\lambda=\tilde{\lambda}=10^6\gg 1$ and $\eta=0$. These parameters are the same as in~\cite{geissler2017}, allowing for comparison of square and triangular lattices. Values of the hopping amplitude and the detuning are varied.

\begin{figure}[pt!]
\includegraphics[width=0.8\linewidth]{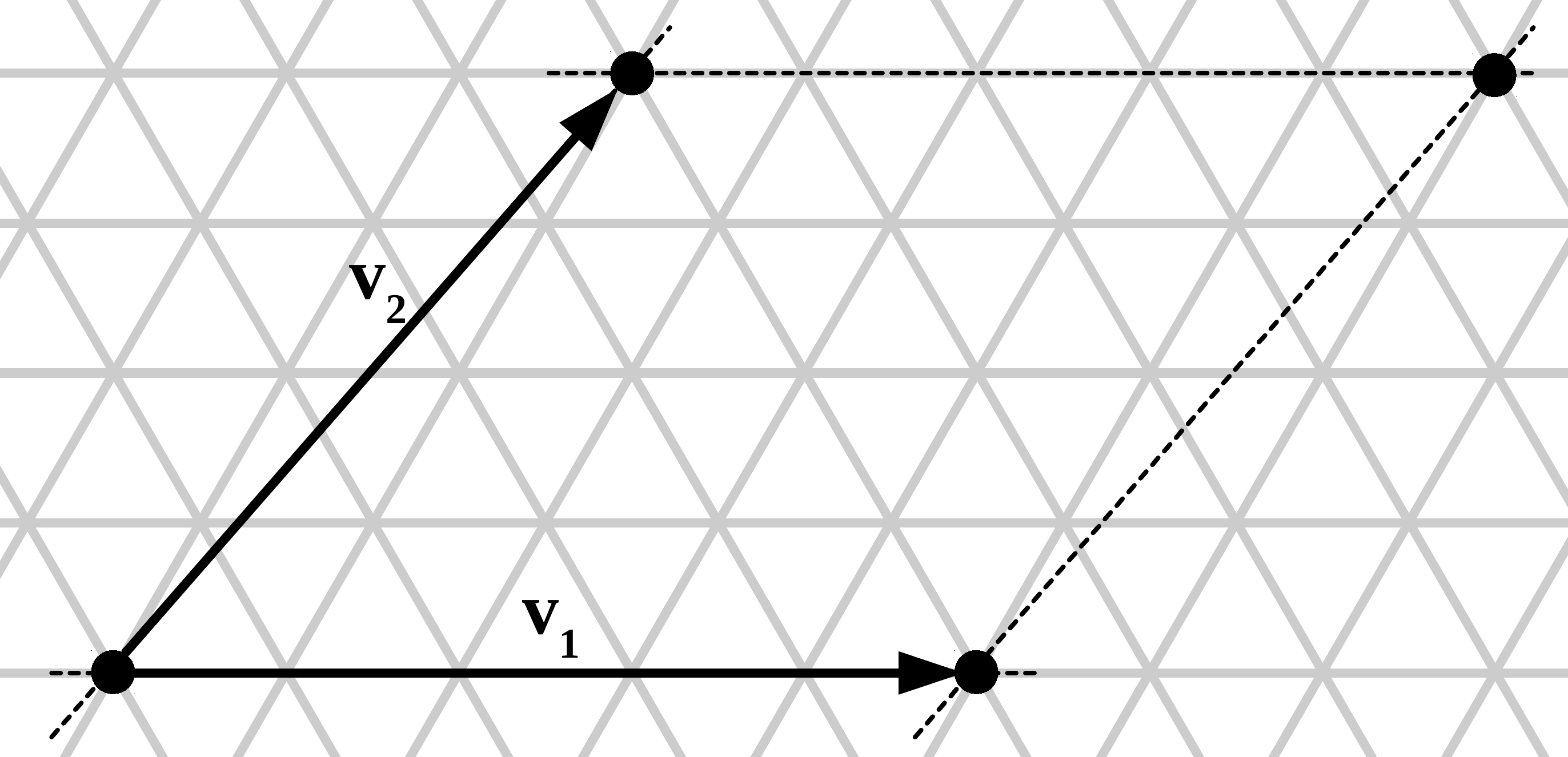}
\includegraphics[width=0.9\linewidth]{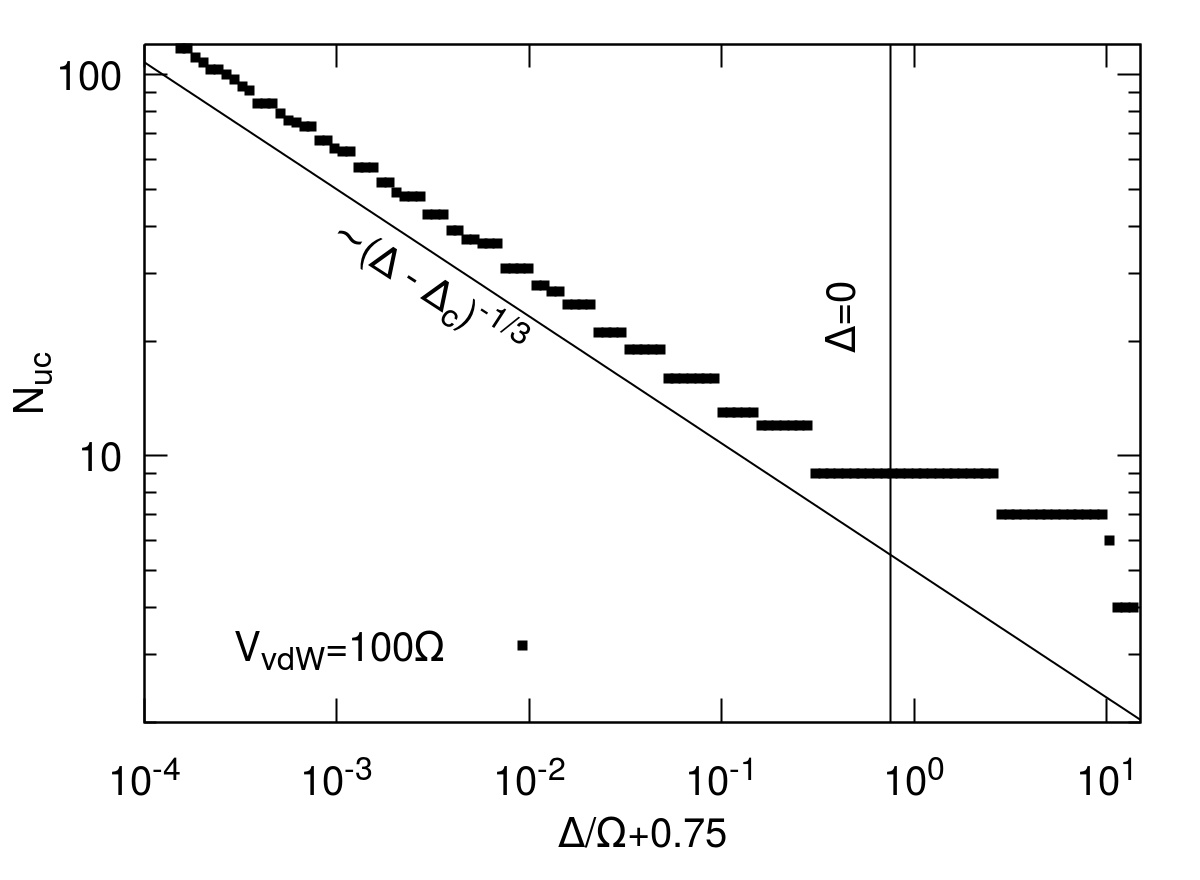}
\caption{\label{fig:frozen} (Top) In the frozen-limit occupied sites are related to each other by a translation by linear combinations of $\mathbf{v}_1$ and $\mathbf{v}_2$. (Bottom) In the frozen-limit changing the detuning results in a series of phase transitions, between different insulating ordered structures characterized by the number of sites per unit cell $N_{uc}$. Parameters are set to: $V_{vdW}=100$, $U=0.1$, $\mu=-0.25$, $J=0$.}
\end{figure}

We first investigate the frozen-limit case $J=0$. In Fig.~\ref{fig:frozen} (bottom) we show how the size of the unit cell of the ground state, given by its number of lattice sites $N_{uc}$, depends on the detuning. Below a critical value of $\Delta_c=-0.75$ the system is empty. As we increase $\Delta$ above $\Delta_c$ we observe a series of phase transitions between insulating ordered (density-wave) phases, resembling the devil's staircase observed in the Ising model~\cite{bak1982,lauer2012,rademaker2013}. Each has a different translational symmetry and size of the unit cell. For values of $\Delta$ close to $\Delta_c$ the unit cell is large, resulting in a very dilute system. As the value of the detuning is increased the density of bosons also increases. These results are qualitatively similar to the ones obtained for a square lattice~\cite{geissler2017}. The trend line close to $\Delta_c$ follows $N_{uc}\sim(\Delta-\Delta_c)^{-\frac{1}{3}}$. The exponent is determined by the spatial dimensionality of the system divided by the exponent in the interaction potential, cf. App.~\ref{app:scaling}.

\begin{figure}[t]
\begin{center}
\includegraphics[width=\linewidth]{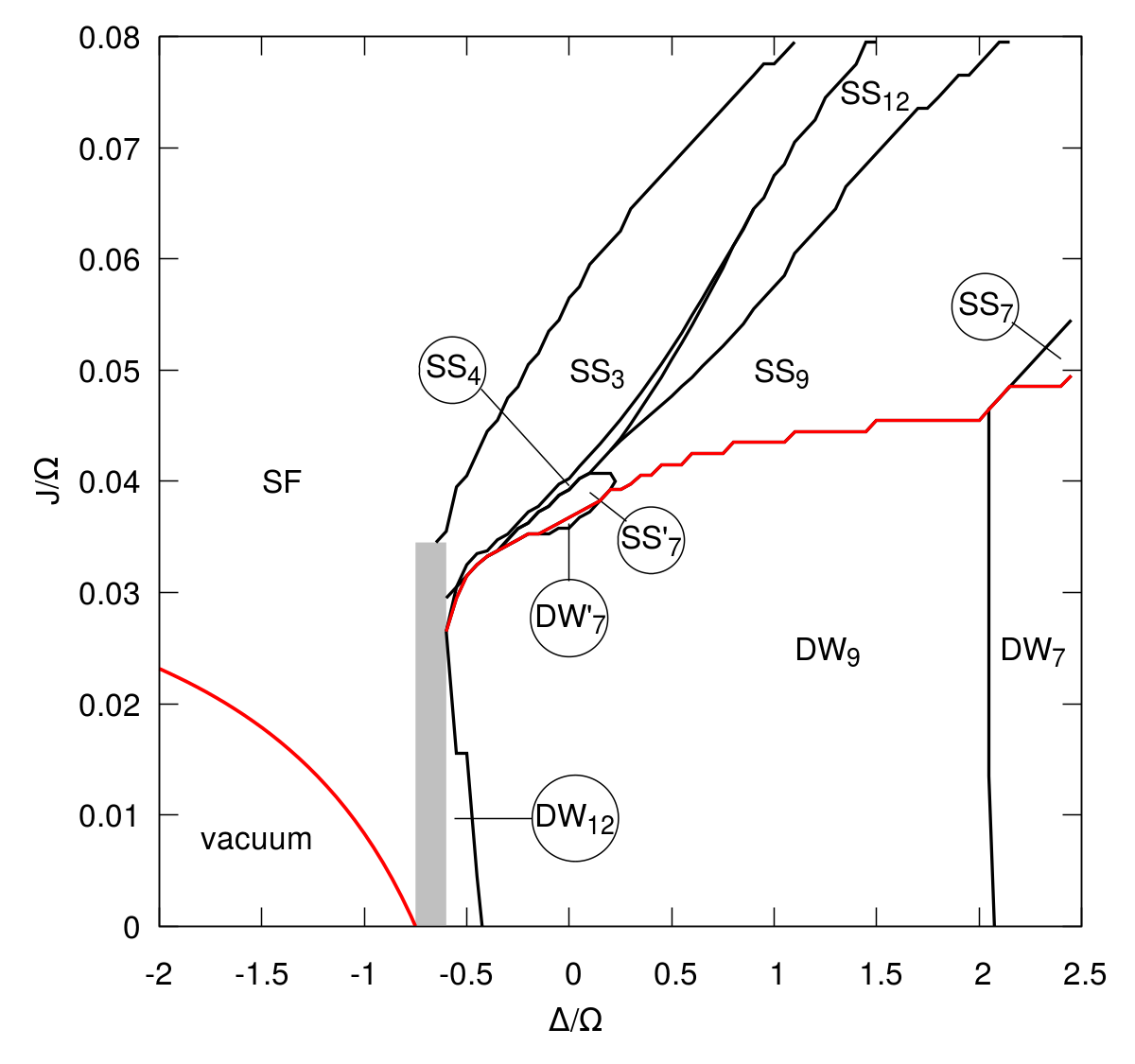}
\caption{\label{fig:phdiag_BDMFT} Phase diagram of the system described by the Hamiltonian~\eqref{eq:fullH} obtained from B-DMFT. Parameters of the system are the same as in Fig.~\ref{fig:frozen} except for the variable hopping amplitude $J$. Phases shown in Fig.~\ref{fig:struct} are labeled $\mathrm{SS}_n$ and $\mathrm{DW}_n$ where the lower index $n=N_{uc}$ represents the number of sites in a unit cell, cf. Tab.\ref{tab:structs}. Gray shading represents an area where B-DMFT calculations did not converge. The red line separates phases breaking the $U(1)$ symmetry from those where it is preserved. For $\Delta<-0.75$ we use formula~\eqref{eq:vacuum} as it matches accurately the B-DMFT data.}
\end{center}
\end{figure}

Next we proceed to the discussion of the finite hopping $J>0$ case within the B-DMFT calculations. Out of a large set of unit-cells considered in the frozen-limit we have selected only the few smallest, relevant for the vicinity of $\Delta=0$, due to the computational complexity of B-DMFT calculations. They are listed in Tab.~\ref{tab:structs}. According to the frozen-limit results the structures that were left out become relevant only in the narrow region of detuning $-0.75<\Delta<-0.7$~\footnote{For much higher detuning one also expects a structure with $N_{uc}=3$ which was not included in Tab.~\ref{tab:structs}. However, as it is a substructure of the one with $N_{uc}=9$ it can still be observed, similarly as $S_6$ in Fig.~\ref{fig:phdiag_BDMFT}.}. For other values of $\Delta$ the structures from Tab.~\ref{tab:structs} should be sufficient.
  
\begin{table}[t!]
\begin{tabular}{c|c|c|c|c}
$\mathbf{v}_1$ & $(2,-1)_{\mathbf{e}}$ & $(2,0)_{\mathbf{e}}$ & $(3,-1)_{\mathbf{e}}$ & $(3,-1)_{\mathbf{e}}$ \\
\hline
 $\mathbf{v}_2$ & $(1,1)_{\mathbf{e}}$ & $(0,2)_{\mathbf{e}}$ & $(0,2)_{\mathbf{e}}$ & $(1,2)_{\mathbf{e}}$ \\
\hline
$N_{uc}$ & $3$ & $4$ & $6$ & $7$ \\
\hline\hline
$\mathbf{v}_1$ & $(3,0)_{\mathbf{e}}$ & $(4,-2)_{\mathbf{e}}$ & $(4,-1)_{\mathbf{e}}$ & $(4,0)_{\mathbf{e}}$  \\
\hline
$\mathbf{v}_2$ & $(0,3)_{\mathbf{e}}$ & $(2,2)_{\mathbf{e}}$ & $(1,3)_{\mathbf{e}}$ & $(0,4)_{\mathbf{e}}$ \\
\hline
$N_{uc}$ & $9$ & $12$ & $13$ & $16$ \\
\end{tabular}
\caption{Spanning vectors $\mathbf{v}_1$ and $\mathbf{v}_2$ and number of sites in the unit cell $N_{uc}$ of the structures considered in the B-DMFT calculations (except for the first one with $N_{uc}=3$, which was not considered explicitly but rather implicitly as a special case of the one with $N_{uc}=9$). $\mathbf{v}_1$ and $\mathbf{v}_2$ are given in the basis of primitive vectors of a triangular lattice $\mathbf{e}_1$ and $\mathbf{e}_2$, e.g. $(3,-1)_\mathbf{e} = 3\mathbf{e}_1 - \mathbf{e}_2$. The primitive vectors in Euclidean space in units of the lattice spacing $a$ are $\mathbf{e}_1 = (1,0)$ and $\mathbf{e}_2 = (\frac{1}{2},\frac{\sqrt{3}}{2})$.}
\label{tab:structs}
\end{table}

\begin{figure*}[ht]
\includegraphics[width=0.245\linewidth]{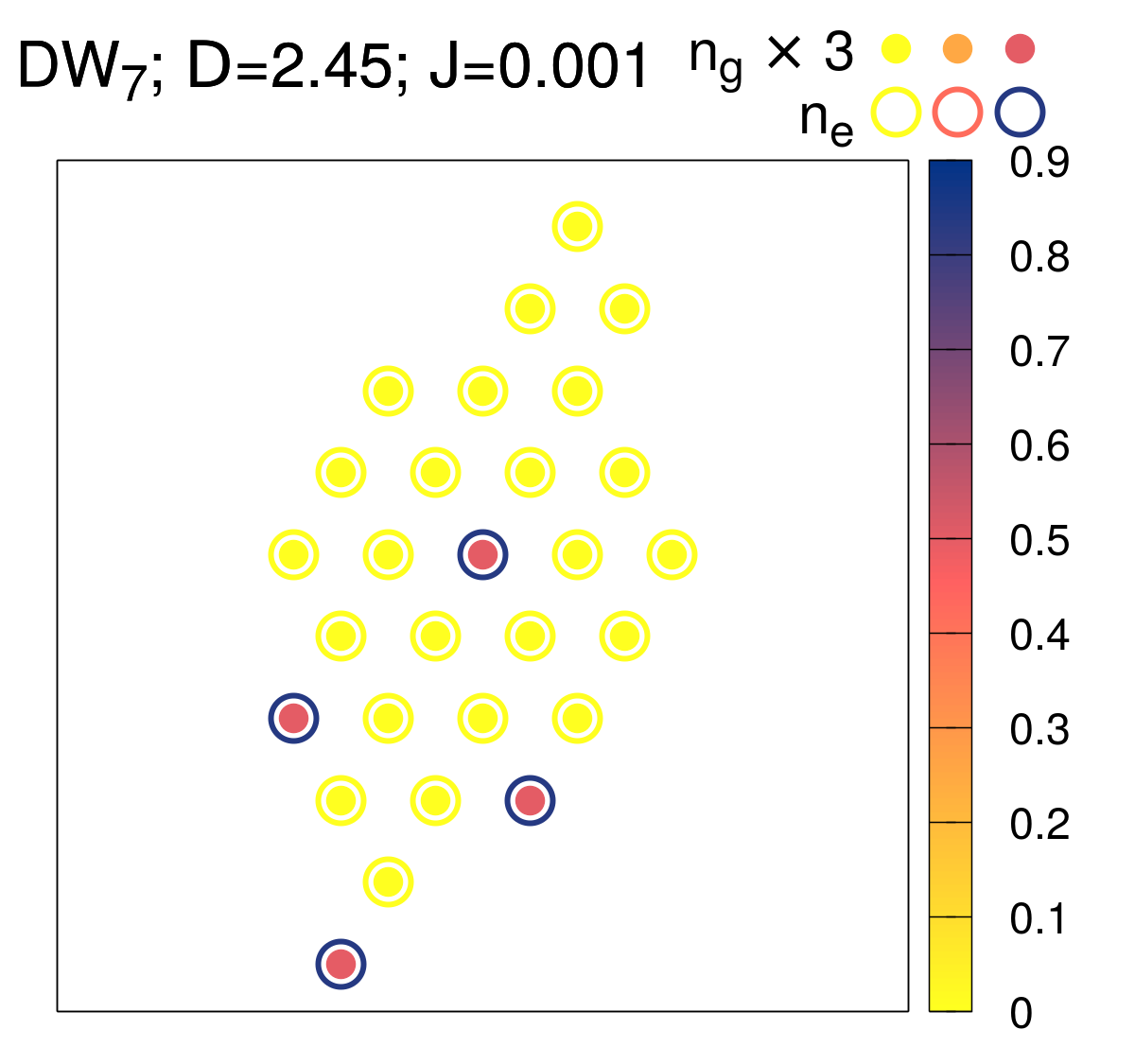}
\includegraphics[width=0.245\linewidth]{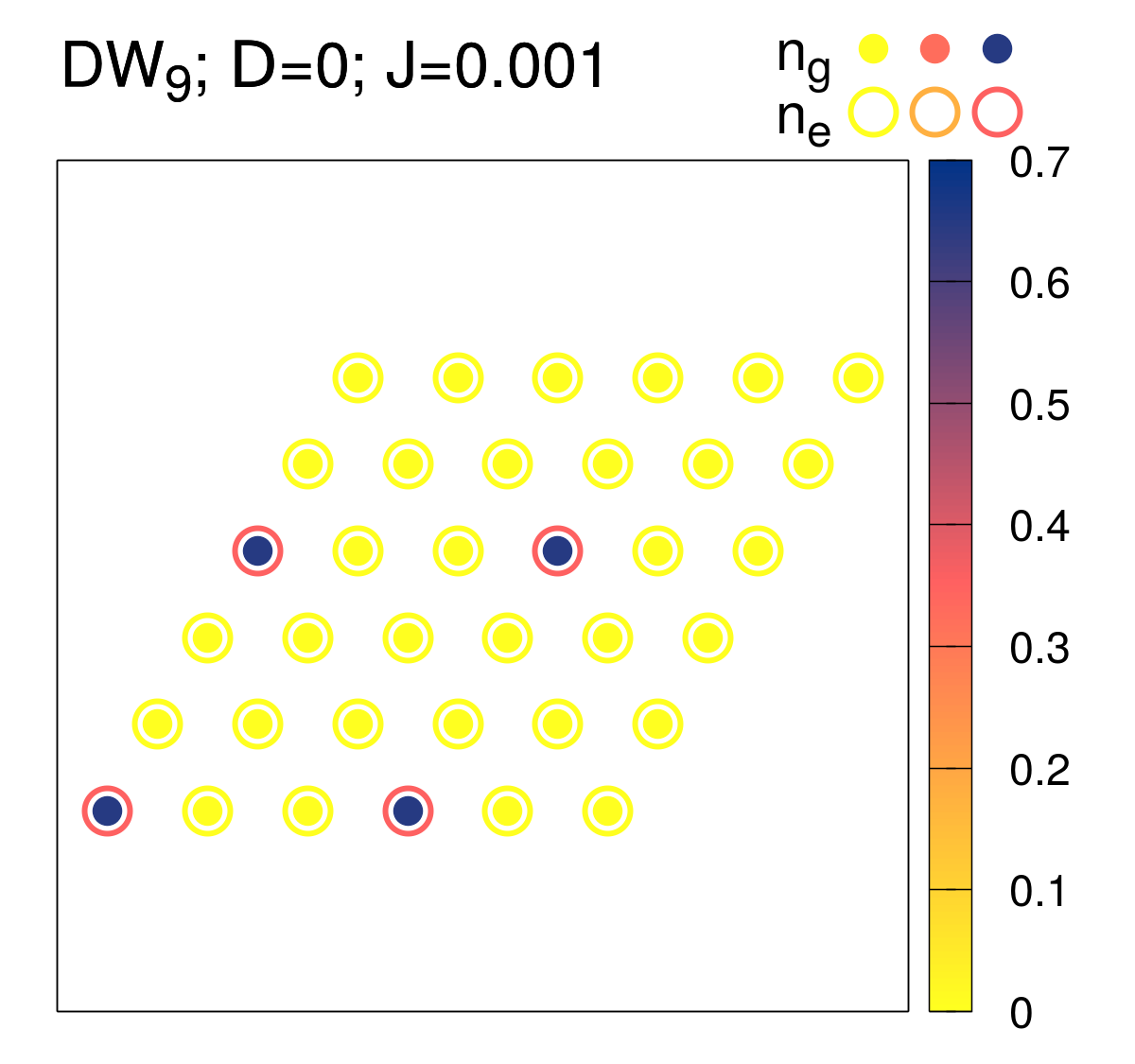}
\includegraphics[width=0.245\linewidth]{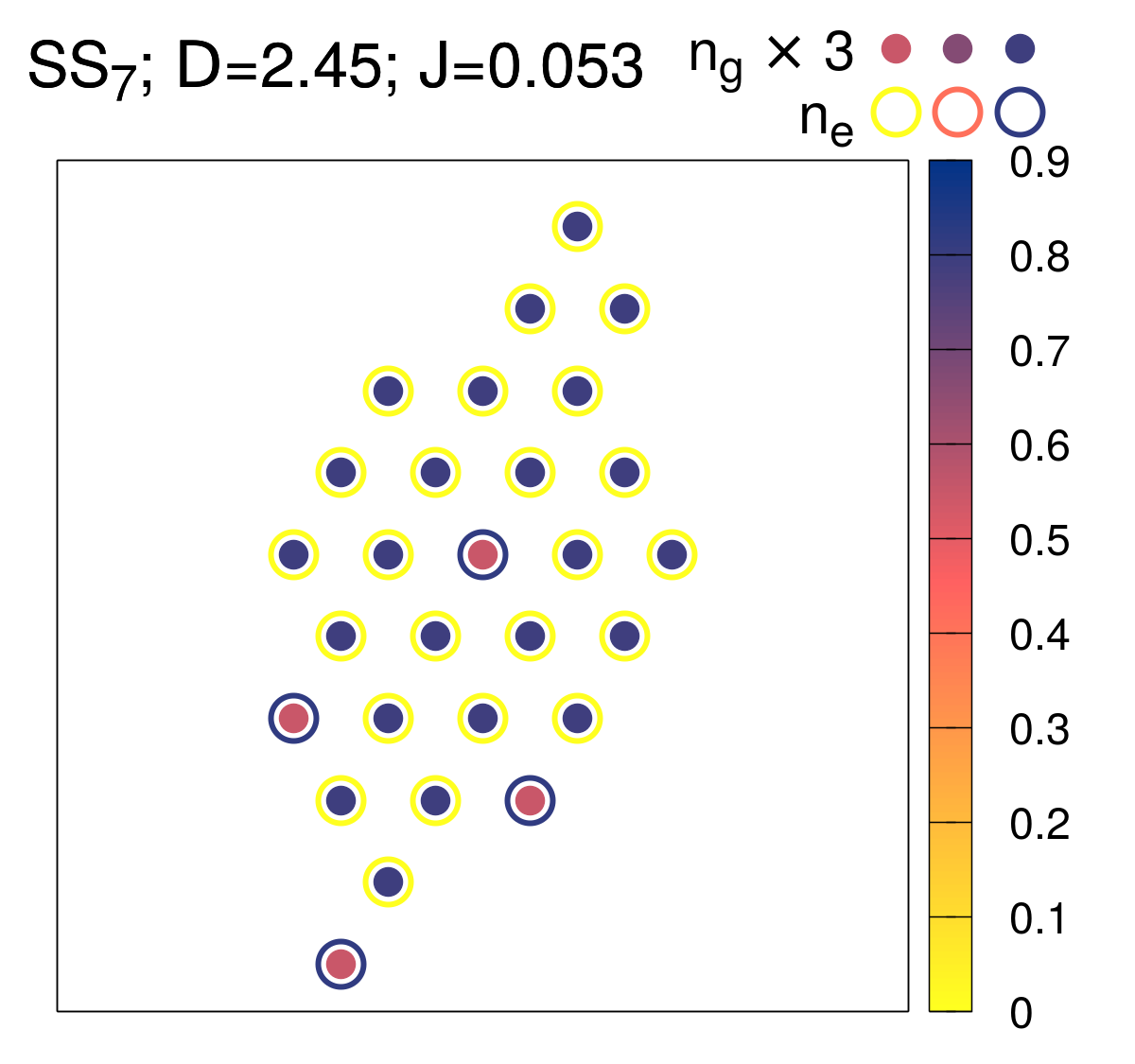}
\includegraphics[width=0.245\linewidth]{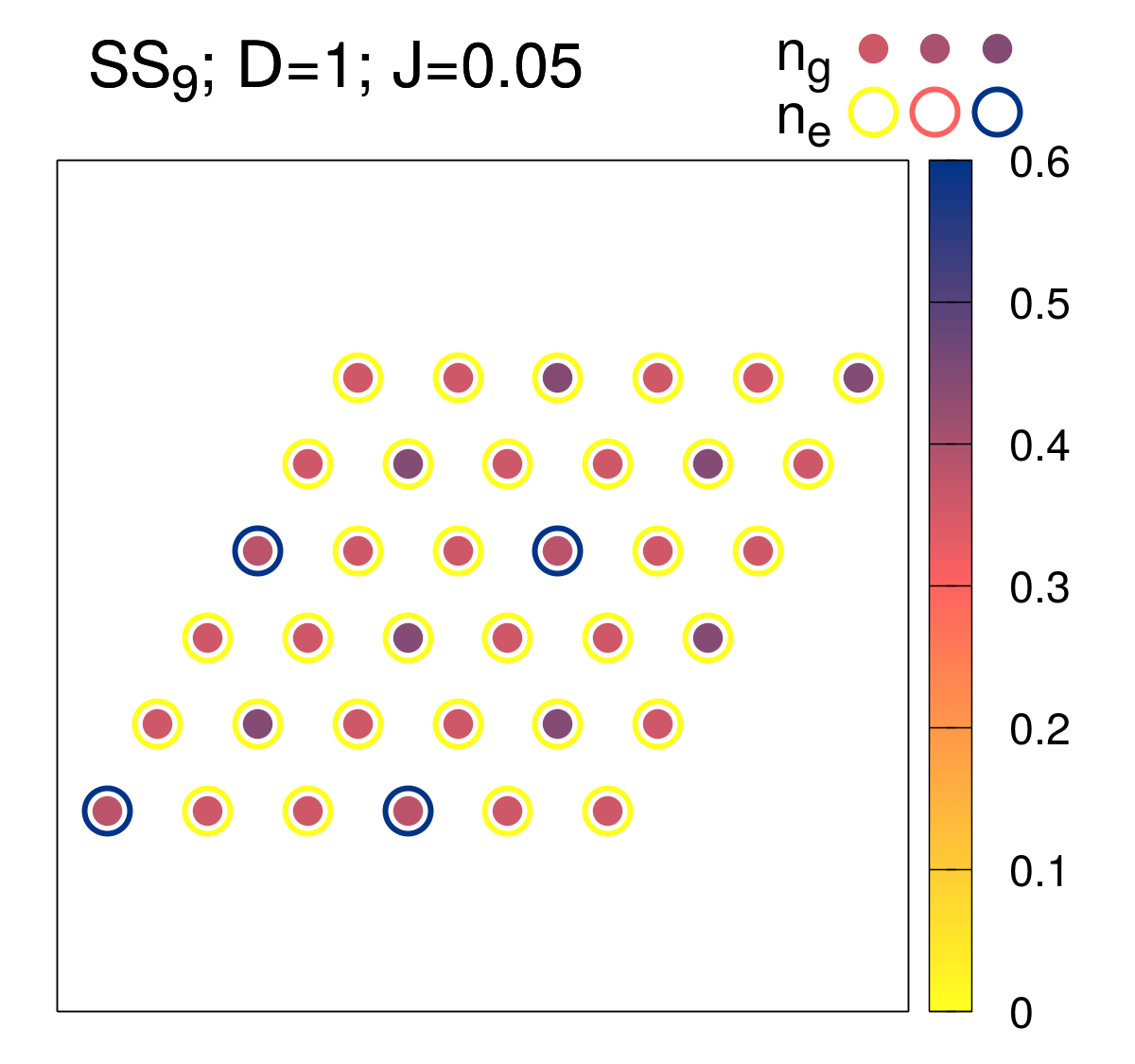}
\includegraphics[width=0.245\linewidth]{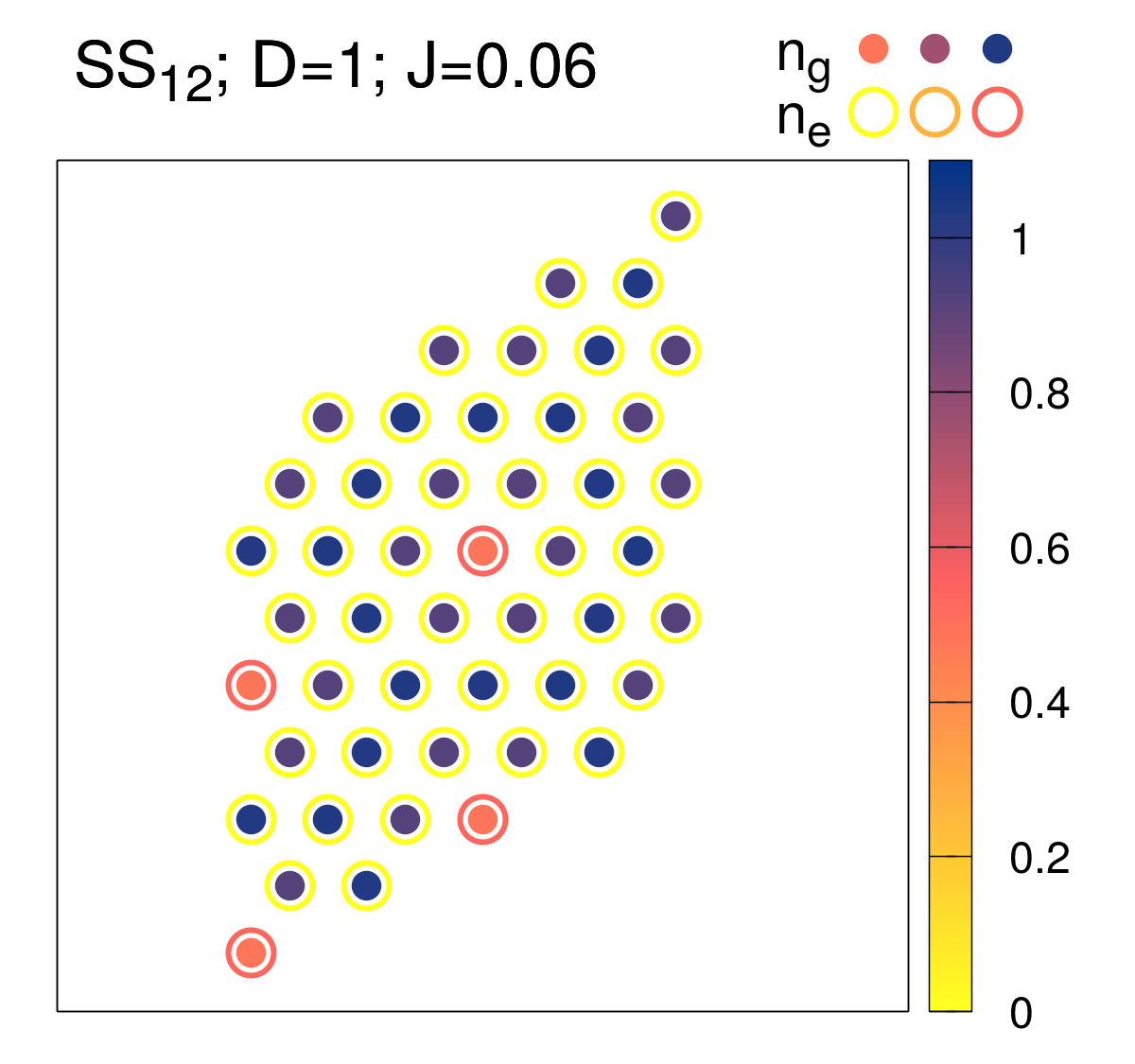}
\includegraphics[width=0.245\linewidth]{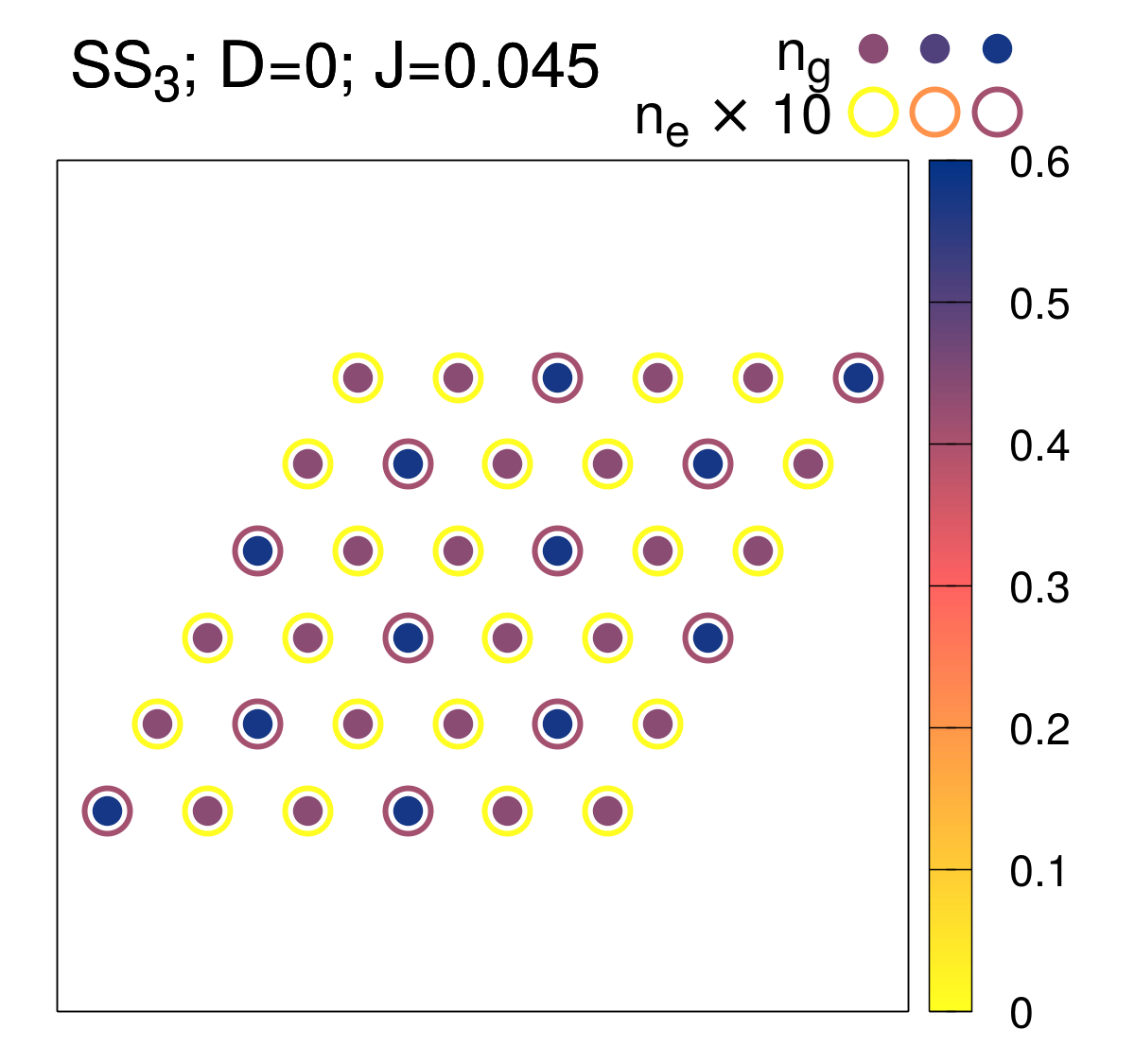}
\includegraphics[width=0.245\linewidth]{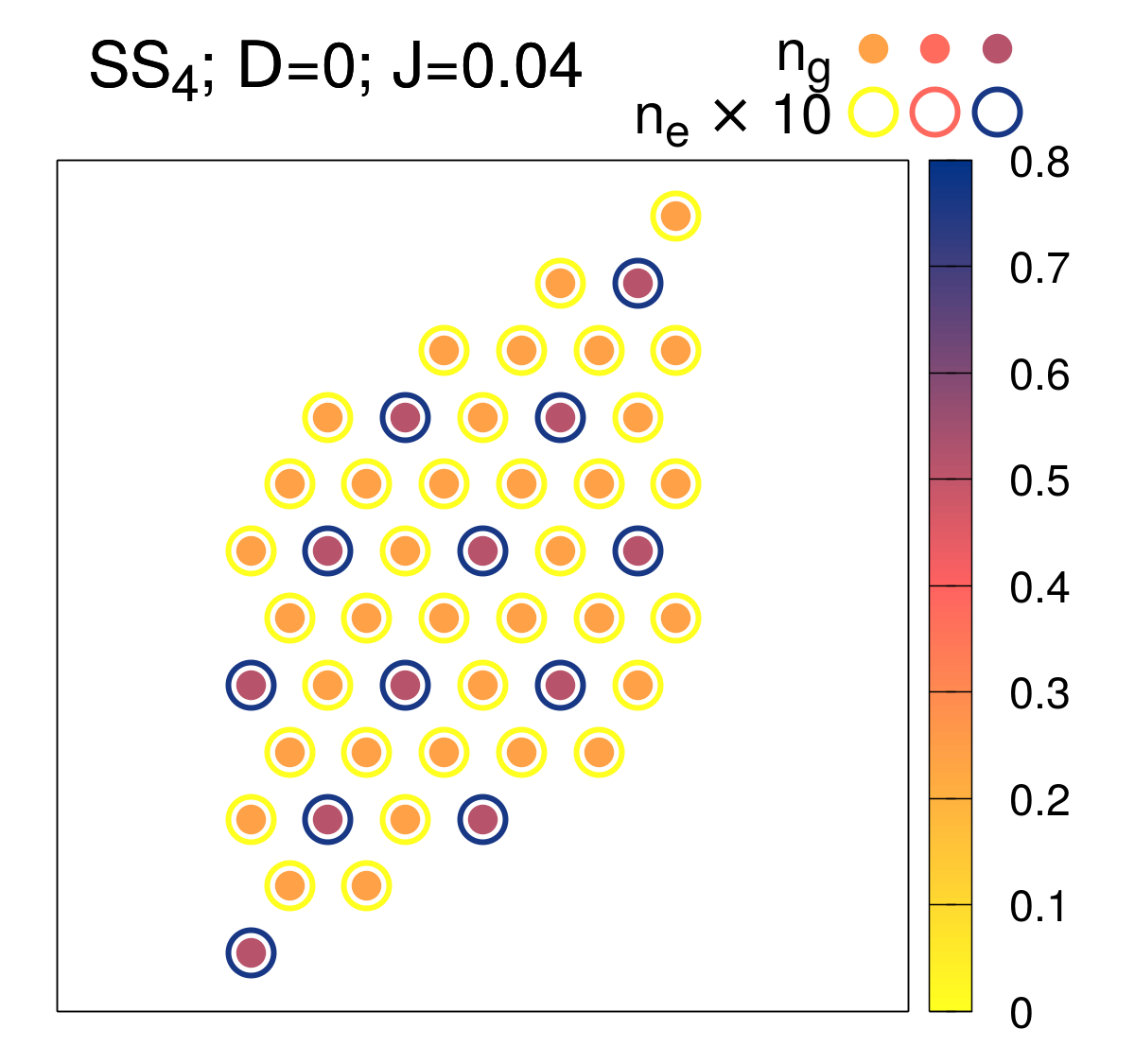}
\includegraphics[width=0.245\linewidth]{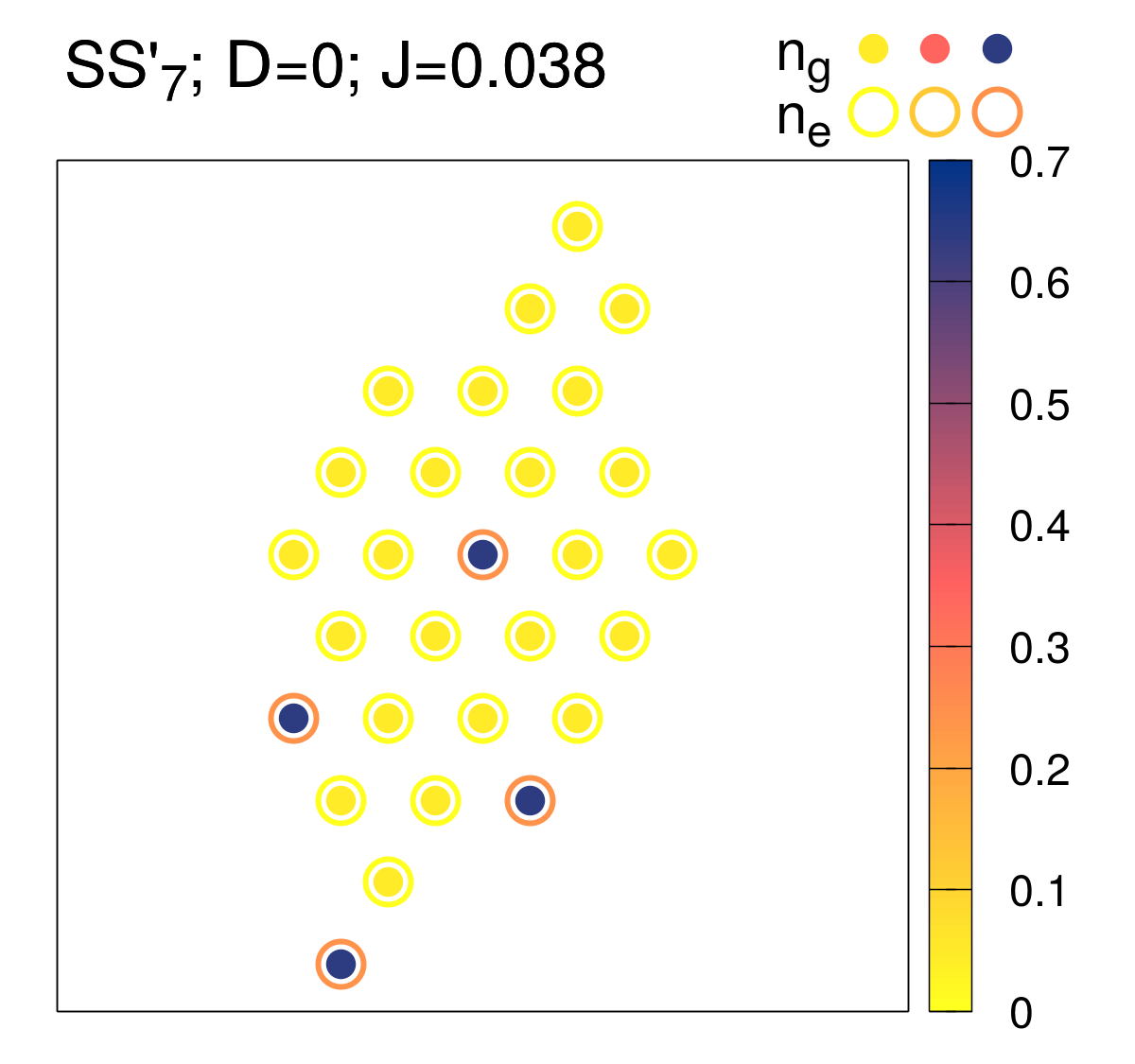}
\caption{\label{fig:struct} Structures of selected phases observed in Fig.~\ref{fig:phdiag_BDMFT}. Color represents the fraction of ground (filled circles) and excited state (empty circles) bosons per lattice site. Each graph represents sites within a quadrupled unit cell of an initial guess. Each site in a unit cell defines a different impurity problem in the B-DMFT procedure. The depicted patterns recur periodically in the lattice. The fractions of excited (ground) state bosons have been rescaled in certain cases for better visibility.}
\end{figure*}

The phase diagram obtained within B-DMFT is shown in Fig.~\ref{fig:phdiag_BDMFT}, while the density-wave patterns observed in different phases are shown in Fig.~\ref{fig:struct}. The phases labeled as $\mathrm{DW}_7$, $\mathrm{DW}'_7$, $\mathrm{DW}_9$ and $\mathrm{DW}_{12}$ are insulating while $\mathrm{SS}_3$, $\mathrm{SS}_4$, $\mathrm{SS}_7$, $\mathrm{SS}'_7$, $\mathrm{SS}_9$, $\mathrm{SS}_{12}$ are supersolid. Lower index indicates number of sites in the unit cell $N_{uc}$ of the structure, cf. Tab.\ref{tab:structs}. In the limit of small hopping amplitude $J=0.001$ we recover the results of the frozen-limit, as expected. The observed structures $\mathrm{DW}_7$ and $\mathrm{DW}_9$ (and $\mathrm{DW}_{12}$, not shown in Fig.~\ref{fig:struct}) follow the trend presented in Fig.~\ref{fig:frozen}. Only the sites of a sublattice defined by vectors $\mathbf{v}_1$ and $\mathbf{v}_2$ are occupied. On each of its sites there is a single boson, which is in a superposition between ground and excited state. The remaining sites of the lattice are nearly empty.

Next we consider the effect of increasing hopping amplitude. For $\Delta<\Delta_c=-0.75$ increasing $J$ leads to a phase transition from the vacuum to a homogeneous superfluid phase. The phase boundary obtained with B-DMFT agrees well with the expression~\eqref{eq:vacuum}. We note that below $\Delta_c$ unlike for the square lattice geometry, where a checkerboard supersolid was found~\cite{geissler2017}, the triangular system does not exhibit any supersolid phase. For $\Delta>-0.6$ small values of the hopping amplitude have only a minor influence on the insulating phases, resulting in small shifts of the phase boundaries with increasing $J$. Further increase of $J$ eventually leads to a spontaneous breaking of the $U(1)$ symmetry in the system and a transition into one of many supersolid phases.

We observe that at higher values of $\Delta$ the supersolid phases immediately above their insulating counterparts have similar order, cf. Fig.~\ref{fig:struct}, $\mathrm{DW}_7-\mathrm{SS}_7$ and $\mathrm{DW}_9-\mathrm{SS}_9$. In these phases, namely $\mathrm{SS}_7$ and $\mathrm{SS}_9$, a significant (when compared to ground state population) number of excited state bosons is present. We observe that both the local ground state condensate order parameter $\langle\hat{a}_i\rangle$ as well as local fluctuations of the occupation $\delta n_i^2=\langle( \hat{n}_{g,i} + \hat{n}_{e,i})^2 \rangle-\langle  \hat{n}_{g,i} + \hat{n}_{e,i} \rangle^2$ (see App.\ref{app:supplementary results}) are much smaller on the sites with excited state bosons than on the surrounding sites with small excited state occupation~\footnote{The condensate of the excited state bosons is small throughout the system, with total fraction an order of magnitude smaller than that of the ground state bosons. Interestingly, even close to the $\mathrm{DW}_9$ phase it shows ordering later observed in $\mathrm{SS}_3$ supersolid.}. We interpret this phase as a supersolid which consists of (i) frozen bosons being in a superposition of ground and excited state (with significant fraction of both) residing in the periodic sublattice of the original lattice, (ii) delocalized, condensed bosons, predominantly of the ground state nature, flowing without friction in the remaining lattice sites. This interpretation is also consistent with the shape of the supersolid--density-wave phase boundary for larger detunings. For a given ordered structure the delocalized bosons can slightly reduce their energy by a small admixture of the excited state, thus increasing the supersolid regime in the phase diagram. This energy reduction is however diminished with increasing fraction of excited state bosons in the ``frozen'' sites, due to strong non-local interaction forces. Hence with increasing $\Delta$ the phase boundary is shifted to higher values of $J$ as the fraction of excited bosons increases.

The supersolid phase $\mathrm{SS}_{12}$ appears to be of the same nature as $\mathrm{SS}_{7}$ and $\mathrm{SS}_{9}$ but with a different spatial structure. The $\mathrm{DW}_{12}$ phase (not depicted in Fig.~\ref{fig:struct}), a $U(1)$ symmetric counterpart of the $\mathrm{SS}_{12}$ phase, appears for small $J$ around $\Delta\approx-0.5$, cf. Fig.~\ref{fig:phdiag_BDMFT}. However, $\mathrm{SS}_{12}$ and its insulating counterpart are not immediately connected in the phase diagram due to the emergence of other phases discussed further in text. 

Starting from the $\mathrm{SS}_{9}$ or $\mathrm{SS}_{12}$ phases and decreasing the detuning $\Delta$ or increasing the hopping amplitude $J$, we observe further phase transitions, cf. Fig.~\ref{fig:phdiag_BDMFT}. Qualitatively different supersolids emerge, labeled as $\mathrm{SS}_{3}$ and $\mathrm{SS}_{4}$. These are characterized by the following features. The fraction of the excited state bosons is significantly (by approximately an order of magnitude) lower than in $\mathrm{SS}_{7}$, $\mathrm{SS}_9$ or $\mathrm{SS}_{12}$. The wavelength of the density-wave pattern is significantly smaller, with smaller distances between sites with non-vanishing fraction of excited state bosons. On these sites we have also observed an increase in condensate fraction and local fluctuations of the occupation, cf. App.~\ref{app:supplementary results}. Finally, we observe a larger density of atoms in the remaining, intermediate sites and more uniform distribution of the condensate order parameter. This behavior with decreasing values of $\Delta$ is opposite of what one would expect if one tried to apply here intuition gained from the frozen-limit. Because of these differences we conclude that this must be a qualitatively different type of supersolid, where we can no longer apply the interpretation of ``frozen'' sublattice sites occupied by the excited state bosons coexisting with condensed ground state bosons in between. These phases bear some resemblance to bubble supersolids~\cite{henkel2012,cinti2014} in that the condensation originates form the sites with non-vanishing excited state fraction, in contrast to the $\mathrm{SS}_7$, $\mathrm{SS}_9$ and $\mathrm{SS}_{12}$ phases. We also suspect that the supersolids observed in $\mathrm{SS}_{3}$ and $\mathrm{SS}_{4}$ could be connected with the concept of defectons~\cite{andreev1969,cinti2014}, which is a condensation of defects (holes) in the ordered structure that are tunneling between different sites. This interpretation seems to be consistent with the observed features of the phases: (i) larger local fluctuations on the sublattice could be related to the presence and condensation of defectons, (ii) smaller wavelength of the observed pattern supports tunneling of defectons in opposition to the larger wavelength patterns, which are more favorable for the condensation of the ground state bosons on the intermediate sites (between the site of the sublattice). However, these arguments alone are not sufficient to confirm this interpretation unambiguously.

It is worth mentioning that there is a relation between $\mathrm{SS}_9$ and $\mathrm{SS}_3$ phases, as well as between $\mathrm{SS}_{12}$ and $\mathrm{SS}_4$ phases. In both cases the symmetry of the former phase can be viewed as a reduced version of the symmetry of the latter phase with $N_{uc}$ reduced by a factor of 3, cf. Fig.~\ref{fig:struct}. However, the $\mathrm{SS}_{12}$ and $\mathrm{SS}_4$ phases, between which we observed a first order phase transition, separate the $\mathrm{SS}_9$ and $\mathrm{SS}_3$ phases from each other.

Regarding the two types of supersolids described above we note that similar observations were made in~\cite{li2018}. There also two types were found, with one consisting of a supersolid of bare (ground state) species and crystalline phase of dressed (coupled to excited state) species. However, due to a different model the second type of supersolid in~\cite{li2018} is not of the same nature as observed here -- one does not observe a significant reduction of the wavelength of the periodic structure but rather an increase.

We also observe a phase at intermediate values of the hopping amplitude $J\approx0.038$ and in the vicinity of detuning $\Delta=0$ which is depicted in Fig.~\ref{fig:struct} and labeled $\mathrm{SS}'_{7}$. It has similar properties to the $\mathrm{SS}_3$ and $\mathrm{SS}_4$ phases, which have a smaller wavelength than the phases at larger detuning, e.g. $\mathrm{SS}_9$, and for which condensation originates from the sites with non-vanishing excited state fraction, cf. App.~\ref{app:supplementary results}. The distinguishing features of $\mathrm{SS}'_7$ are that the fraction of excited state bosons is significantly higher than in $\mathrm{SS}_3$ or $\mathrm{SS}_4$ and that we also observed an insulating counterpart of the $\mathrm{SS}'_7$ phase, namely the $\mathrm{DW}'_7$ phase. Both do not occur in the mean-field calculations, as discussed in the next section.

We note that the results presented here are qualitatively similar to the ones presented in~\cite{geissler2017} for a square lattice. A significant difference is visible here only near the boundary between superfluid and supersolid phases. On a square lattice this boundary separates superfluid from checkerboard supersolid, which extends to values of detuning below $\Delta_c$, where one can induce a phase transition from superfluid to supersolid by increasing (rather than decreasing) the hopping amplitude. On a triangular lattice the checkerboard supersolid cannot exist due to frustration and no supersolid phase exists below $\Delta_c$. Apart from this the two phase diagrams are similar. We attribute this to the fact that the spacing (wavelength) in the majority of structures observed here is larger than the lattice spacing. In this case the difference in geometry of the two lattices has a weaker impact. For these longer-wavelength structures it is actually the square lattice that becomes more frustrated with respect to the favored (due to van der Waals interaction) Wigner crystal formation~\cite{rademaker2013} than the triangular lattice.

\subsection{Comparison with the static mean-field}
\label{sec:sub2sec3}

\begin{figure}[pt!]
\begin{center}
\resizebox{1.0\columnwidth}{!}{
\includegraphics{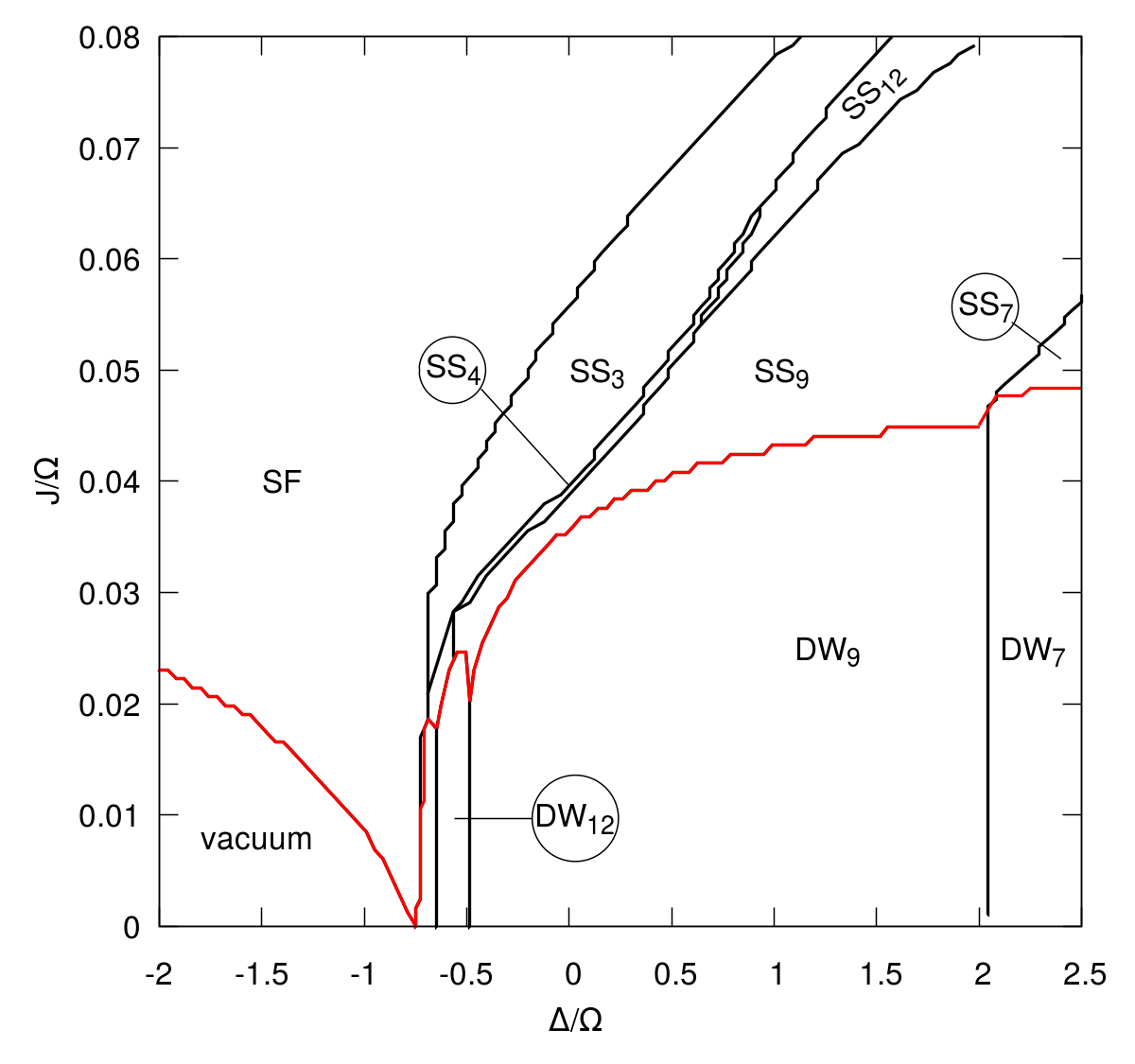}}
\caption{\label{fig:mean_field} Phase diagram obtained within Gutzwiller (static) mean-field approach. The parameters are the same as in Fig.~\ref{fig:phdiag_BDMFT}.}
\end{center}
\end{figure}%
In Fig.~\ref{fig:mean_field} we present the results of calculations performed within the static Gutzwiller mean-field approximation~\cite{fisher1989,barbier2018}. Upon comparison with the B-DMFT we notice that both methods give similar results. The main features of the phase diagrams agree well. Below we focus on the most relevant differences.

As the static mean-field approach favors ordered phases, we expect a phase transition between insulating and superfluid (supersolid) phases to apprear at lower values of the hopping amplitude. Indeed, comparing Fig.~\ref{fig:phdiag_BDMFT} with Fig.~\ref{fig:mean_field} we observe that the boundary of the insulating phases is shifted downwards. This effect is almost negligible for larger absolute values of the detuning and becomes relevant only in the region $|\Delta|\lessapprox 0.75$. This is also a region where an increasing number of phases compete in the system. It seems that only in this region local quantum fluctuations will significantly affect the system's behavior.

Another discrepancy arises from the oversimplification of the insulating phases within the static mean-field approach. The boundaries between different phases with $U(1)$ symmetry do not depend on the hopping amplitude and their positions are uniquely defined by the detuning. In contrast, B-DMFT calculations show that finite hopping can induce a phase transition between two density-wave phases.

The above two observations are directly related to the most significant difference that we observed. Namely, in the static mean-field neither the supersolid $\mathrm{SS}'_7$ nor the density-wave $\mathrm{DW}'_7$ phase was observed, which should be present according to B-DMFT, cf. Fig.~\ref{fig:phdiag_BDMFT}. This occurs in the region, where we observed largest discrepancies between the two methods in the values of the condensate order parameter and in properties of the insulating density-wave phase. We conclude that as the detuning $\Delta$ approaches critical value $\Delta_c=-0.75$ at intermediate values of the hopping amplitude, effects of local quantum fluctuations become significant (this is further backed up by comparing different self-consistency conditions, App.~\ref{app:supplementary results}). At this point we emphasize for clarity that non-local fluctuations are treated in both methods on the same level (Hartree mean-field). Therefore, we cannot make definite statements about their significance.

\begin{figure}[pt!]
\begin{center}
\resizebox{1.0\columnwidth}{!}{
\includegraphics{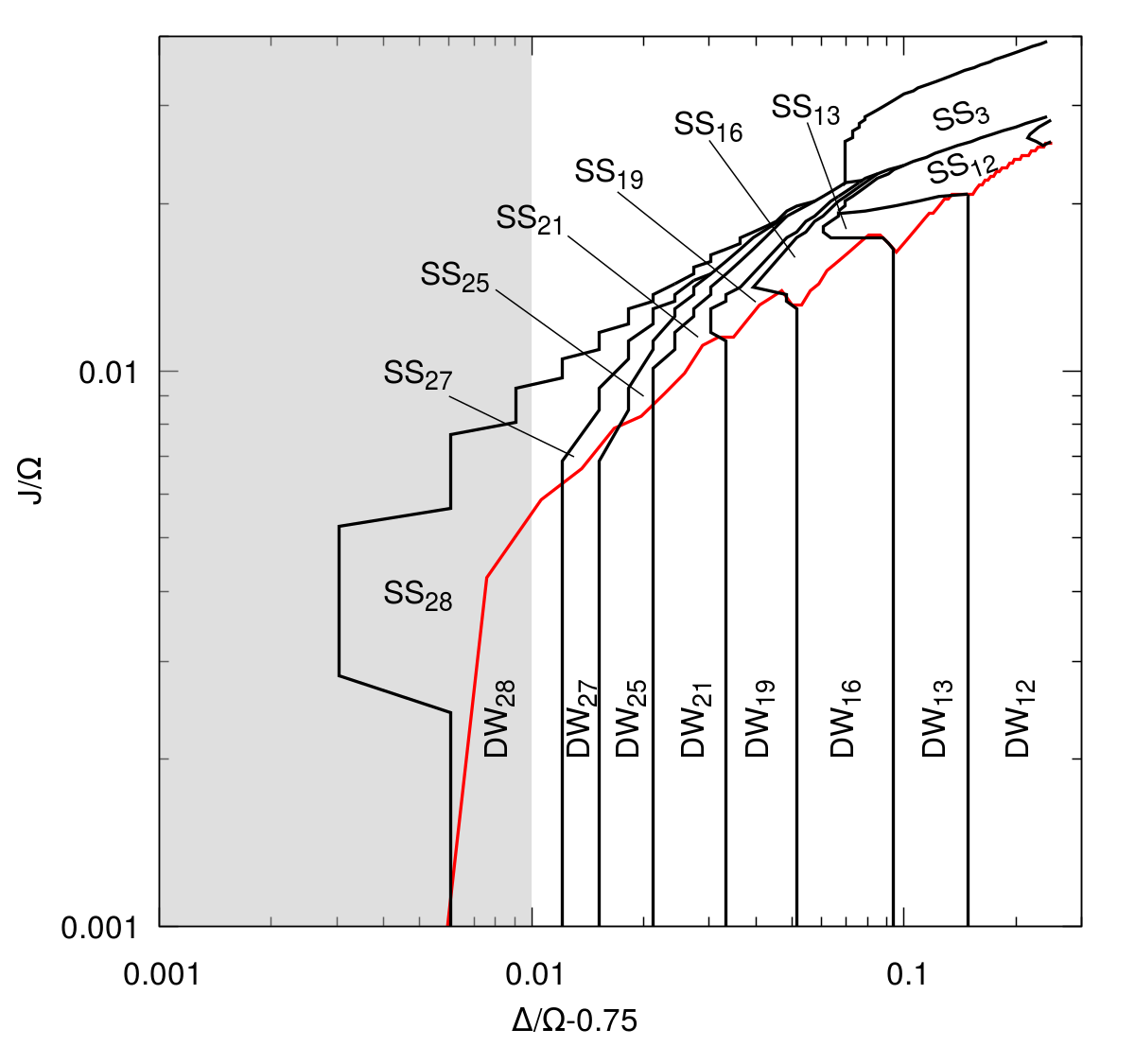}}
\caption{\label{fig:mean_field_crit} Phase diagram obtained within Gutzwiller (static) mean-field approach in the critical region. The parameters are the same as in Fig.~\ref{fig:phdiag_BDMFT}. Gray shading represents the range of parameters beyond the limit of accuracy of the method due to the maximal size of the considered crystalline structure.}
\end{center}
\end{figure}

The last discrepancy between the results of B-DMFT and static mean-field can be observed in the extent of the $\mathrm{SS}_4$ and $\mathrm{SS}_{12}$ phases. In static mean-field it is slightly smaller than in B-DMFT due to larger extent of the $\mathrm{SS}_{9}$ phase. Nevertheless, the remaining features of the phase diagram are qualitatively accurately captured by the static mean field. As this method is significantly less demanding computationally, it is the best that we can do at this stage to get some insight into the critical region of $\Delta\approx\Delta_c$, where it is difficult to obtain converged B-DMFT results. We study this region of the phase diagram in Fig.~\ref{fig:mean_field_crit}.
As we approach $\Delta_c$ from above, we encounter a series of phase transitions forming a devil's staircase (note the logarithmic scale), both of the insulating density-wave phases at small hopping amplitude as well as of the supersolid phases for larger hopping. When investigating the devil's staircase in the supersolids we notice that increasing the hopping amplitude seems to favor longer wavelength structures shifting the devil's staircase pattern to higher detunings. This is similar to what we observed for $\Delta\gtrapprox 1$, cf. Fig.~\ref{fig:mean_field}. However, this trend seems to be reversed at the intermediate detuning of $\Delta\gtrapprox -0.68$ which coincides with the onset of $\mathrm{SS}_{3}$ phase (short wavelength) for larger hopping amplitude. We suspect that this might be a feature emerging due to the competition between the two general types of supersolid discussed in Sec.~\ref{sec:sub1sec3} and could lead to a multicritical point around $\Delta\approx-0.68$ and $J\approx0.022$. However, investigating this region within a more accurate B-DMFT method requires improvements of our implementation of the method and goes beyond the scope of this work.

\subsection{Finite-size system with inhomogeneous Rabi frequency}
\label{sec:sub3sec3}

In Sec.~\ref{sec:sub1sec3} we have established a relation between the behavior of long-range interacting bosons on the triangular and square lattices. In order to make this comparison, we have chosen the same parameters as used in~\cite{geissler2017}. However, these values of the parameters are not optimally suited for experimental realization of the model~\eqref{eq:fullH} with Rydberg atoms loaded into an optical lattice. Relatively large values of the hopping amplitude $J$ and small values of the van der Waals interaction $V_{vdW}$ with respect to the Rabi frequency would require using a Rydberg excited state with low principal quantum number $n\sim 16$. These states have a short lifetime due to spontaneous emission and dephasing processes induced by the black-body radiation~\cite{beterov2009}. Below we propose more realistic parameters, which were chosen based on the observations made in the previous sections and for which a supersolid phase could be observed.

The main two sources of dissipation in the system are the spontaneous emission and black-body induced dephasing~\cite{beterov2009}. The lifetime $\tau$ of an excited Rydberg state depends approximately as $\tau\sim n^3$ on the principal quantum number $n$. It is therefore advantageous to increase $n$.

Another issue one needs to overcome is the avalanche dephasing~\cite{zeiher2017,boulier2017,geissler2018}. A black-body radiation-induced transition of a single atom to another Rydberg state might trigger an avalanche of atom loss from the system. The average time after which such a process occurs is given by~\cite{boulier2017}
\begin{equation}\label{eq:lifetime}
\tau_c = \tau\left( b\sum_i\langle \hat{n}_{e,i}\rangle \right)^{-1},
\end{equation}
where $b$ is the branching ratio of the excited state. $\tau_c$ is inversely proportional to the total number of excited state atoms in the system. Therefore, the best candidate for experimental observation of a supersolid phase is the $\mathrm{SS}_3$ phase where the excited state fraction is very low. However, since in~\eqref{eq:lifetime} the total number of Rydberg atoms in the system appears, rather than their density, we need to consider relatively small system sizes. This leads to the further issue of increased Rydberg fraction at sharp edges of the finite size system, such as shown in~\cite{vermersch2015,geissler2018}. In order to avoid this problem we additionally choose a Gaussian profile of the Rabi laser with a narrow waist on the order of several $\mu$m, as suggested in~\cite{geissler2018}, given by
\begin{equation}\label{eq:profile}
\Omega_i = \Omega \ \mathrm{exp}\left[{\frac{|\mathbf{i}-\mathbf{0}|^2}{\kappa^2}}\right],
\end{equation}
with $\mathbf{0}$ corresponding to the position of the center of the system and $\Omega_i$ an effective Rabi frequency at site $i$.

Taking the above restrictions into account we consider a system with bosonic $^{87}$Rb atoms loaded into a two dimensional triangular optical lattice, e.g. such as described in~\cite{becker2010} with lattice spacing $a\approx 0.5\mu$m. For such a system one should be able to achieve a hopping amplitude of approximately $\sim0.45\hbar$kHz~\cite{becker2010}. The local interaction is a tunable parameter, which we set to $U=1\hbar$kHz. We choose to couple the ground state to an excited $|26S\rangle$ Rydberg state by the Rabi term. Using this $|26S\rangle$ state on a lattice with spacing $a\approx 0.5\mu$m gives a van der Waals interaction strength on the order of $V_{vdW}\approx 1600\hbar$MHz~\cite{singer2005} and mean lifetime of $\tau\approx10\mu$s~\cite{beterov2009}. Rabi frequency and detuning on the order of $\Omega\approx0.1\hbar$MHz, $\Delta=-0.4\hbar$MHz should also be feasible experimentally. The remaining issue is to focus the Rabi laser such that it has a Gaussian profile~\eqref{eq:profile} with $\kappa=3.5\mu$m.

\begin{figure}[pt!]
\begin{center}
\resizebox{1.\columnwidth}{!}{
\includegraphics{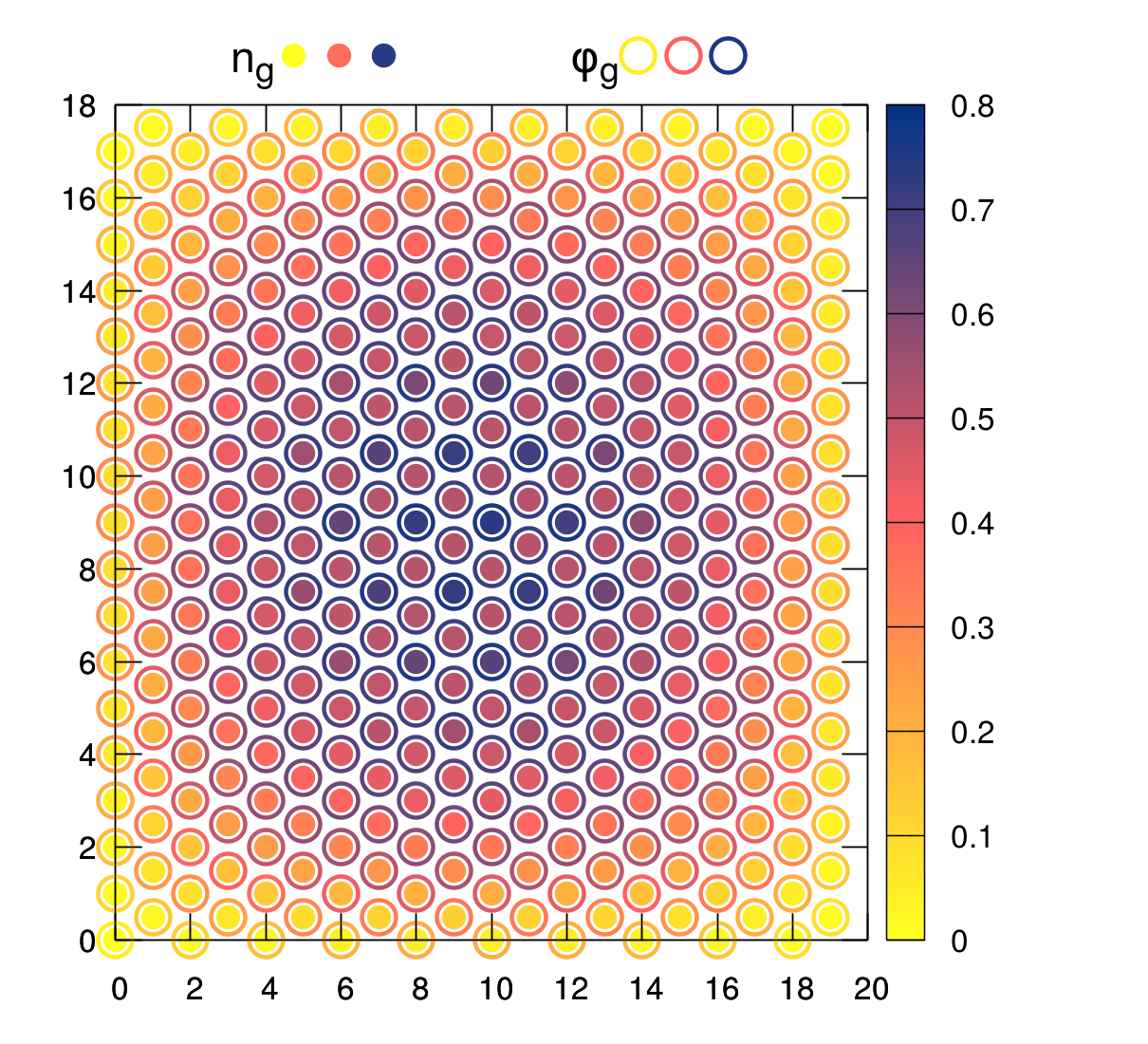}}
\caption{\label{fig:real_space} Ground state $\langle\hat{n}_{g,i}\rangle$ and excited state $\langle\hat{n}_{e,i}\rangle$ occupation in real space for an inhomogeneous Rabi frequency $\Omega_i$ given by~\eqref{eq:profile}. The parameters of the system are: $J=0.0045\Omega$, $\mu=-0.025\Omega$, $U=0.01\Omega$, $\Delta=-4\Omega$, $V_{vdW}=1.6\times 10^4\Omega$.}
\end{center}
\end{figure}

Setting everything in units of $\Omega$ and $a$ we obtain the parameters of the simulation to be the following:
$J=0.0045\Omega$, $\mu=-0.025\Omega$, $U=0.01\Omega$, $\Delta=-4\Omega$, $V_{vdW}=1.6\times 10^4\Omega$ and $\kappa=7a$. The resulting density pattern is shown in Fig.~\ref{fig:real_space}.
In the center of the system we observe the same structure as in the $\mathrm{SS}_3$ phase. Its visibility, defined as
\begin{equation}\label{eq:visibility}
\mathcal{V}=\frac{\langle \hat{n}_{g,i} \rangle - \langle \hat{n}_{g,j} \rangle}{\langle \hat{n}_{g,i} \rangle + \langle \hat{n}_{g,j} \rangle},
\end{equation}
where $i$ corresponds to the site with maximal occupation and $j$ to the nearest-neighbor of $i$, has the value $\mathcal{V}\approx 0.175$.

Lastly we note that the average total number of atoms in the system is $\sum_i \langle \hat{n}_{g,i} + \hat{n}_{e,i}\rangle\approx 113$ while the average total number of Rydberg excitations is $\sum_i \langle \hat{n}_{e,i} \rangle \approx 0.0373 $. Together with the branching ratio on the order of $b\approx0.1$~\cite{boulier2017} the average time after which the avalanche is set off is $\tau_c\approx 3$ms, which is comparable with the characteristic timescale of the hopping process $t\sim\hbar/J$, and thus is promising for experimental realization. The remaining challenges are to achieve low enough temperatures of the system and to determine whether dissipative effects other than avalanche dephasing can destroy the order in the supersolid phase. The latter has been investigated in Ref.~\cite{barbier2018} and it seems that even for relatively large dissipation strength the order prevails for times on the order of hundreds of $\mu$s if the fraction of excited state atoms is small, similarly as in the case of avalanche dephasing.

\section{Conclusions}
\label{sec:end}

In this work we have studied the effect of frustration on the formation of crystalline and supersolid states in the extended Bose-Hubbard model with two bosonic species, one itinerant and one subject to two-body van der Waals long-range interaction. The two species are also coupled by a Rabi term and local interaction. We have focused on a system at zero temperature, without including explicitly the effects of coupling to the environment. We have also used the Hartree mean-field approximation to decouple the long-range interaction term.

Within B-DMFT the same model has been previously investigated on a two-dimensional square lattice~\cite{geissler2017}. Here we solve the problem with two methods: B-DMFT and the Gutzwiller static mean-field approach. Comparison of the B-DMFT results for the two lattice geometries allowed us to determine the effect of the (frustrated) triangular geometry on ordered states. Comparison of B-DMFT results to those of the static mean-field approach allowed us to estimate the significance of local quantum fluctuations.

We have obtained a rich phase diagram, including: insulating density-wave, superfluid and supersolid phases. We observed that the phase diagram on the triangular lattice is qualitatively similar to the one observed for a system with square lattice geometry. Within the parameter regimes considered the only significant discrepancy is the absence of the checkerboard ordered supersolid and of the supersolid below a critical value of the detuning $\Delta_c$ (determined in the frozen-limit). The similarity of the results for the two geometries can be attributed to the low density of atoms and the large wavelength of the observed ordered structures, when compared to the lattice spacing.

Comparison of the B-DMFT and static mean-field methods shows an overall good agreement between the two approaches. The differences are limited to small regions of the phase diagram and small phase-boundary shifts.

We have also studied the model on a finite-size lattice with Gaussian profile of the Rabi term, where we have chosen experimentally convenient parameters. We have found a supersolid phase with a low fraction of excited atoms and a visible spatial modulation of the density. We believe this to be the most promising approach for experimental realization of supersolid phases with Rydberg atoms, minimizing the effects of dissipation.

\begin{acknowledgments}
Support by the Deutsche Forschungsgemeinschaft via DFG SPP 1929 GiRyd and the high-performance computing center LOEWE-CSC is gratefully acknowledged. The authors also acknowledge useful discussions with C. Gro\ss{}, S. Hollerith, W. Li, Y. Li, S. Whitlock and J. Zeiher.
\end{acknowledgments}

\appendix

\section{Scaling of $N_{uc}$ close to $\Delta_c$ in the frozen-limit}
\label{app:scaling}

Below we give a simplified argumentation for the dependence of the size of the unit cell $N_{uc}$ on the detuning close to the critical detuning strength $\Delta_c$ in the frozen-limit $J=0$. We assume the system is $d$-dimensional. We first consider the energy gain due to adding a single particle into an empty system. In such case the only relevant energy scales in the Hamiltonian~\eqref{eq:fullH} are the chemical potential $\mu$, the detuning $\Delta$ and the Rabi frequency $\Omega$. In order to have a finite value of $\Delta_c$, which marks the transition of the system to vacuum, we set $\mu<0$. Finding the single-particle eigenstates in such case is simple and their energies are given by
\begin{equation}\label{appeq:eigen}
e_{\pm} = -\mu -\frac{\Delta\pm\sqrt{\Delta^2+\Omega^2}}{2}.
\end{equation}
The low energy state is given by $e_+$, and when $e_+<0$ it is energetically favorable to put the particles in the system. Thus the condition $e_+=0$ determines the critical value of detuning $\Delta_c$ (e.g. in case of $\Omega=1$ and $\mu=-0.25$ we get $\Delta_c=-0.75$, cf. Sec.~\ref{sec:sub1sec3}). Note that this argumentation can be extended beyond the frozen-limit of $J=0$. By taking a completely delocalized single particle state of the ground state bososns, giving $-zJ$ contribution to the energy, one can obtain critical value of the hopping amplitude $J_c$ as a function of the chemical potential, the detuning and the Rabi frequency, yielding~\eqref{eq:vacuum}. We further consider the energy gain per particle in the vicinity of $\Delta_c$, with $\Delta=\Delta_c+\delta$ and $0<\delta \ll 1$. Expanding~\eqref{appeq:eigen} up to first order in $\delta$ we get
\begin{equation}\label{appeq:eigenex}
e_+ \approx -\frac{1}{2}\left(1+\frac{\Delta_c}{\sqrt{\Delta_c^2+\Omega^2}} \right)\delta.
\end{equation}
Therefore, energy gain per particle due to adding particles in the system is proportional to $\delta=(\Delta-\Delta_c)$.

However, upon adding particles into the system we increase the potential energy due to the van der Waals interaction. We therefore need to estimate the energy cost due to having a certain density of particles in the system. We assume that the particles form a uniform Wigner crystal with certain density $\rho\sim 1/N_{uc}$. In such case the average distance between particles is given by $r_c\sim \rho^{-1/d}$. The energy per particle due to the van der Waals interaction can be estimated by integral
\begin{equation}\label{appeq:envdW}
e_{vdW}\sim \rho\int_{r_c}^{\infty} \frac{r^{d-1}}{r^6} \d r = \rho\frac{r_c^{d-6}}{6-d}\sim \rho^{6/d}.
\end{equation}
If we now require the energy cost of interaction to be compensated by the energy gain due to the Rabi frequency we obtain $(\Delta-\Delta_c)\sim \rho^{6/d} \sim N_{uc}^{-6/d}$, which for $d=2$ gives $N_{uc}\sim(\Delta-\Delta_c)^{-1/3}$, cf. Fig.~\ref{fig:frozen}.

\begin{figure}[b!]
\begin{center}
\resizebox{0.95\columnwidth}{!}{
\includegraphics{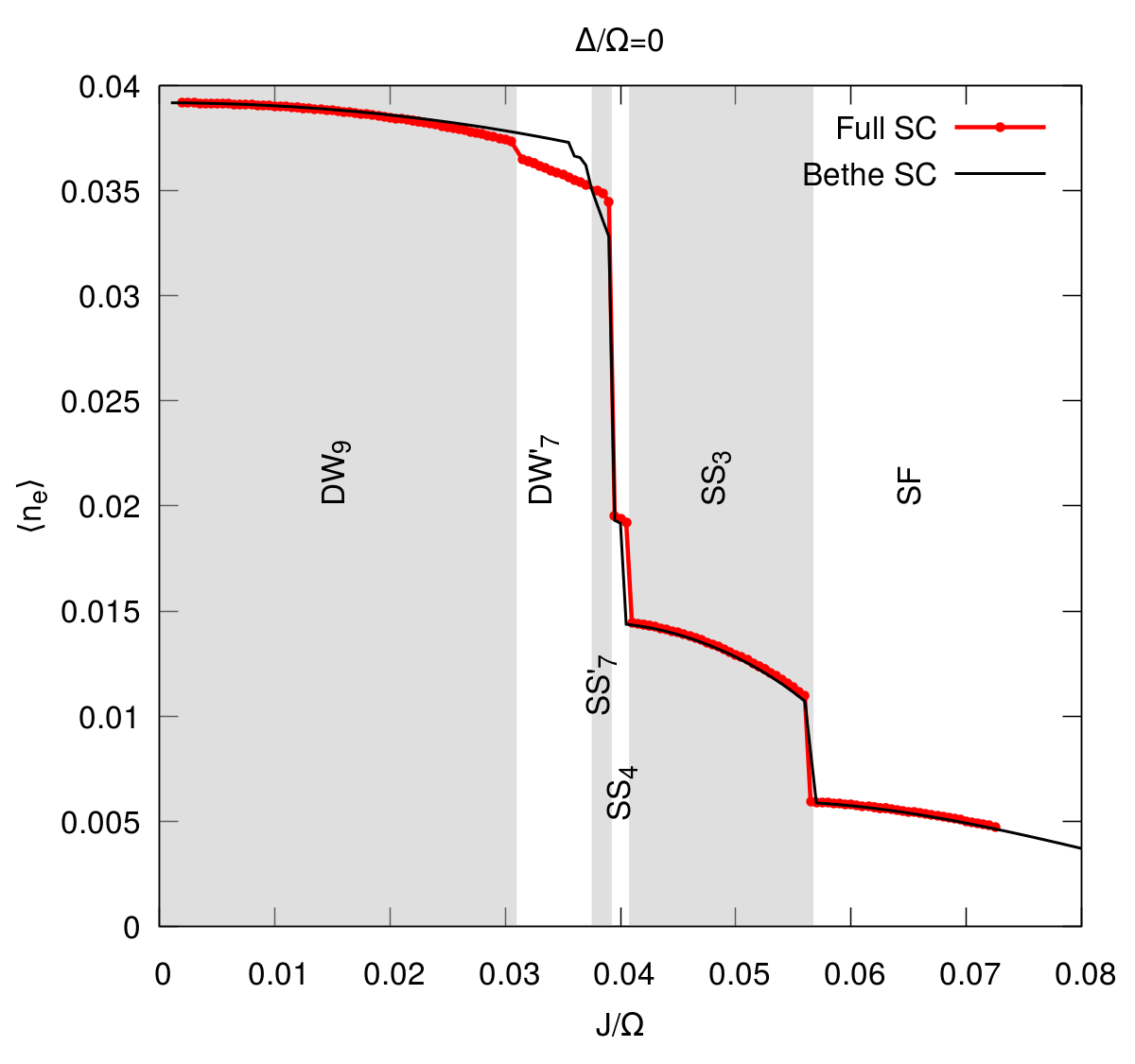}}
\caption{\label{fig:SC_test} Average occupation of the excited state bosons $\langle \hat{n}_e\rangle$ as a function of hopping amplitude for $\Delta=0$. Other parameters are the same as in Fig.~\ref{fig:phdiag_BDMFT}. Red line with points represents results obtained with full B-DMFT self-consistency conditions. Black line represents results obtained with simplified self-consistency condition given by~\eqref{eq:betheSC}. Gray shading represents ranges of different phases labeled in the graph as determined by the full self-consistency.}
\end{center}
\end{figure}%
\begin{figure*}[t!]
\includegraphics[width=0.32\linewidth]{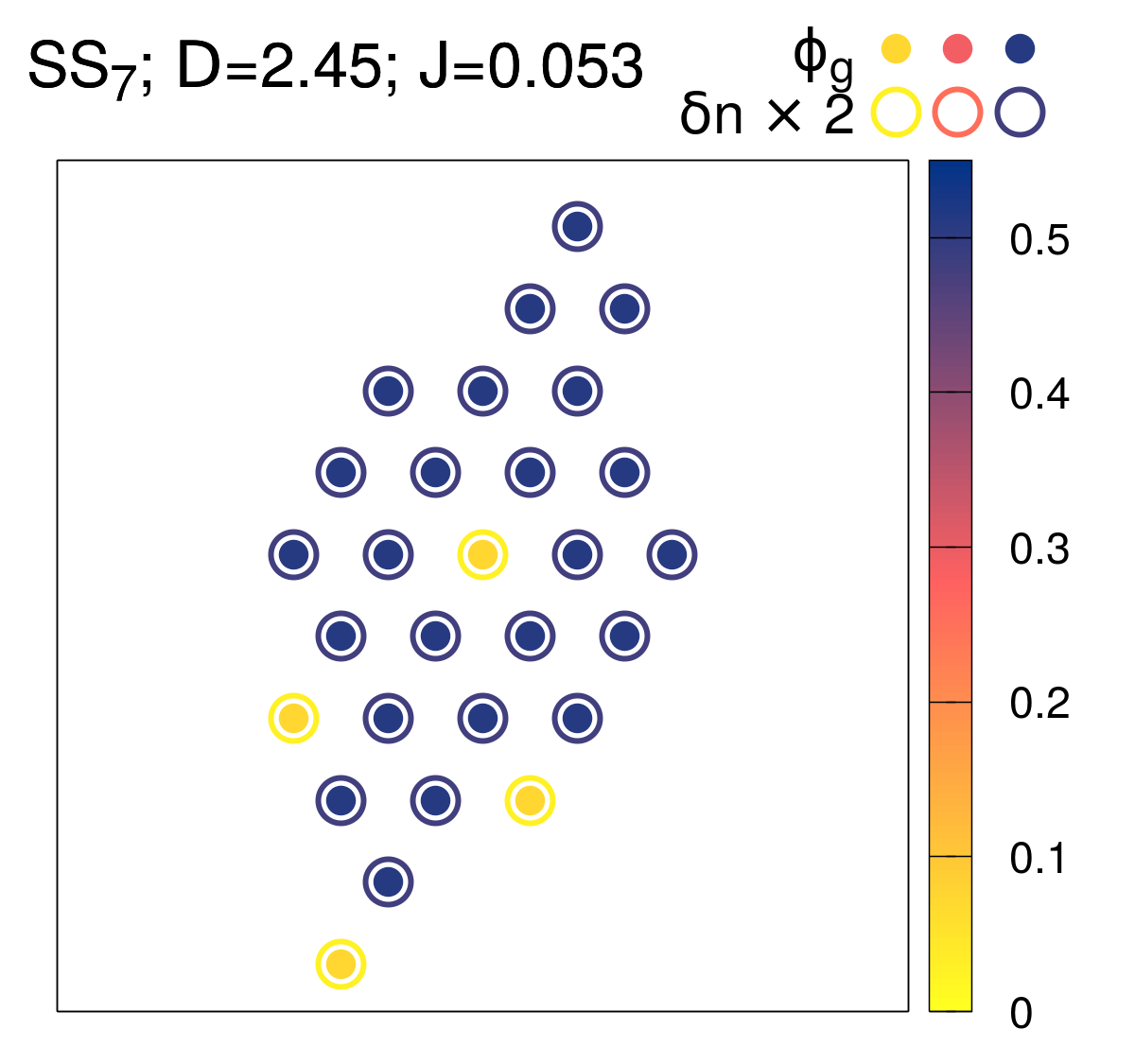}
\includegraphics[width=0.32\linewidth]{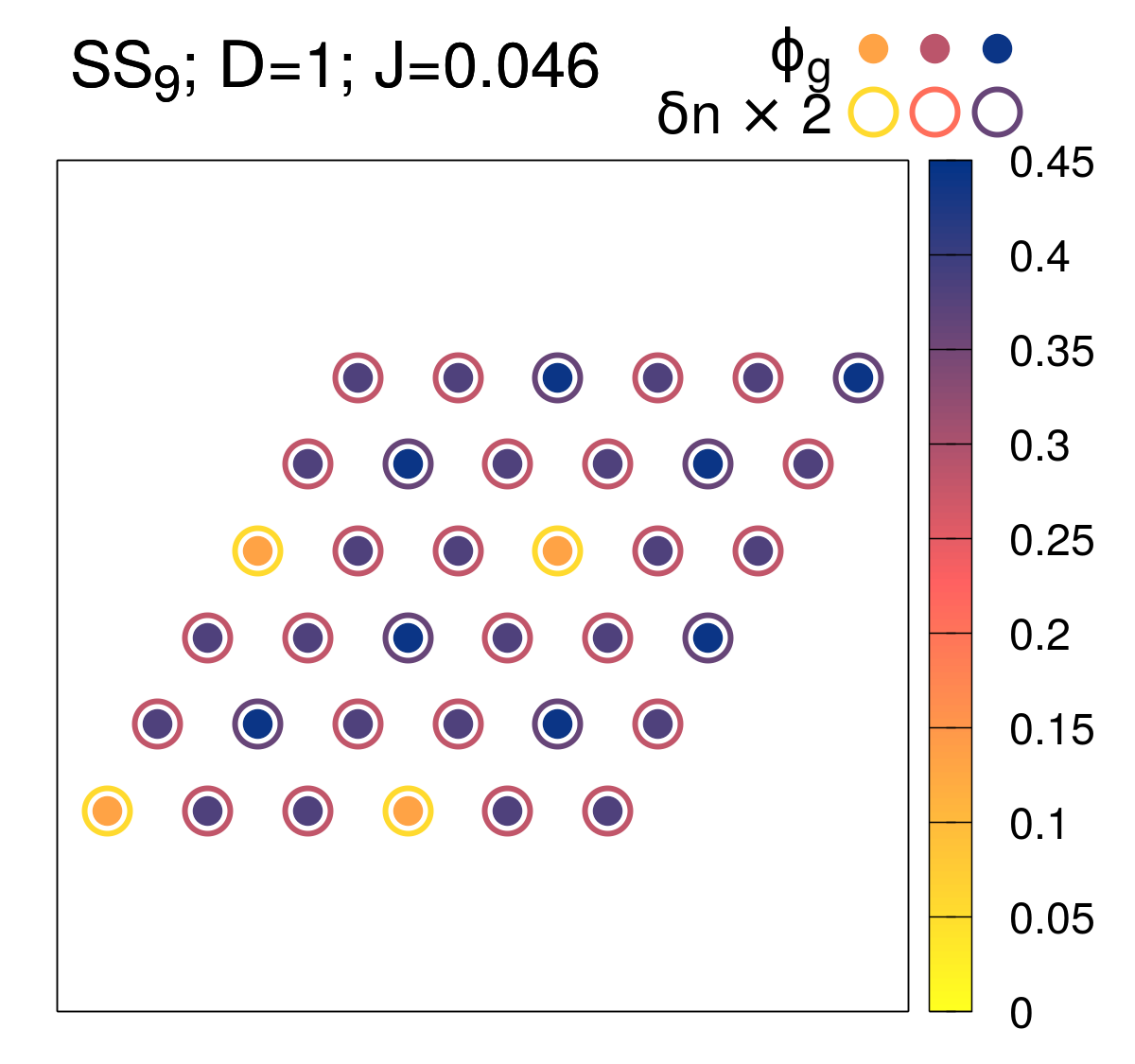}
\includegraphics[width=0.32\linewidth]{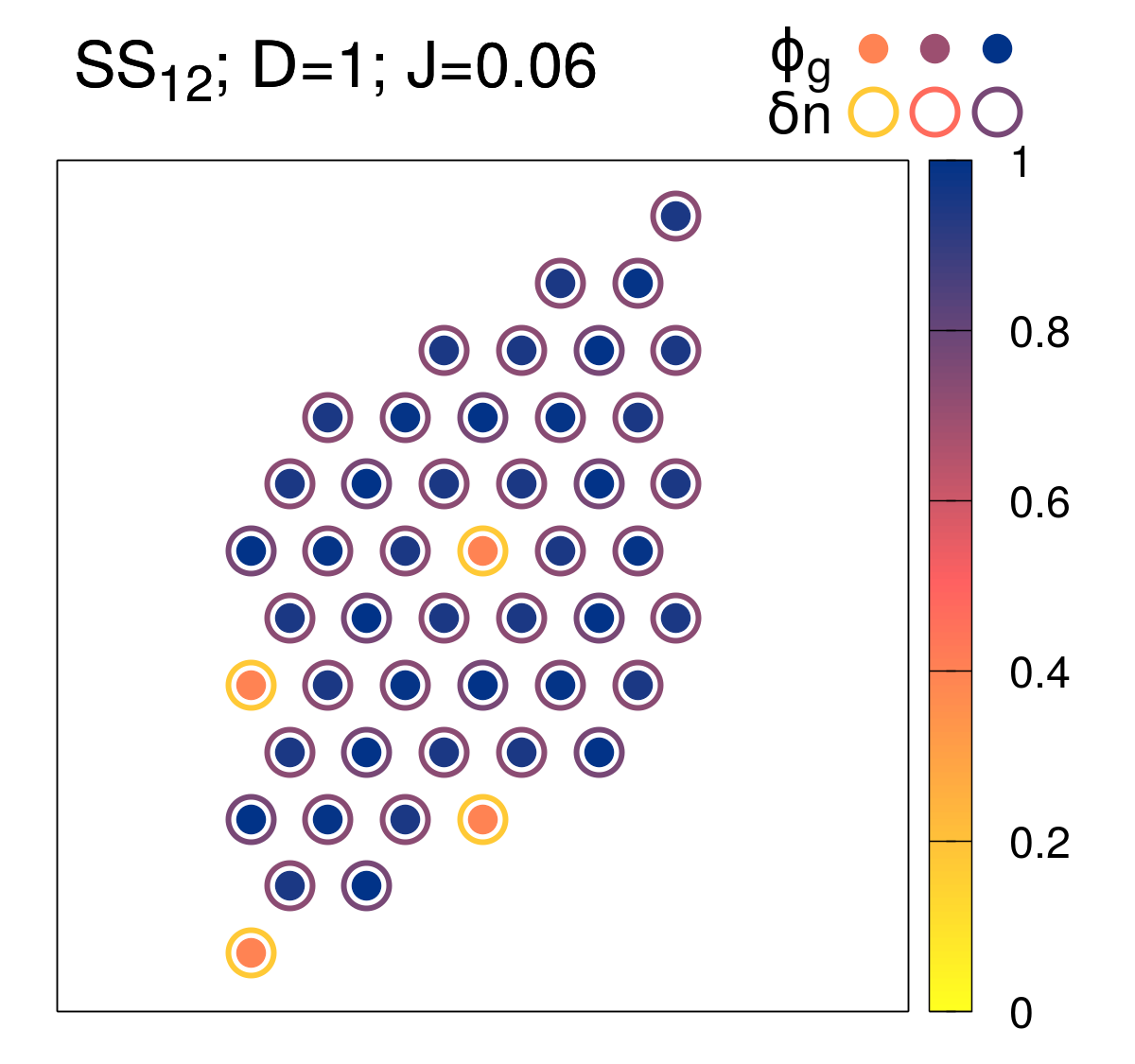}
\includegraphics[width=0.32\linewidth]{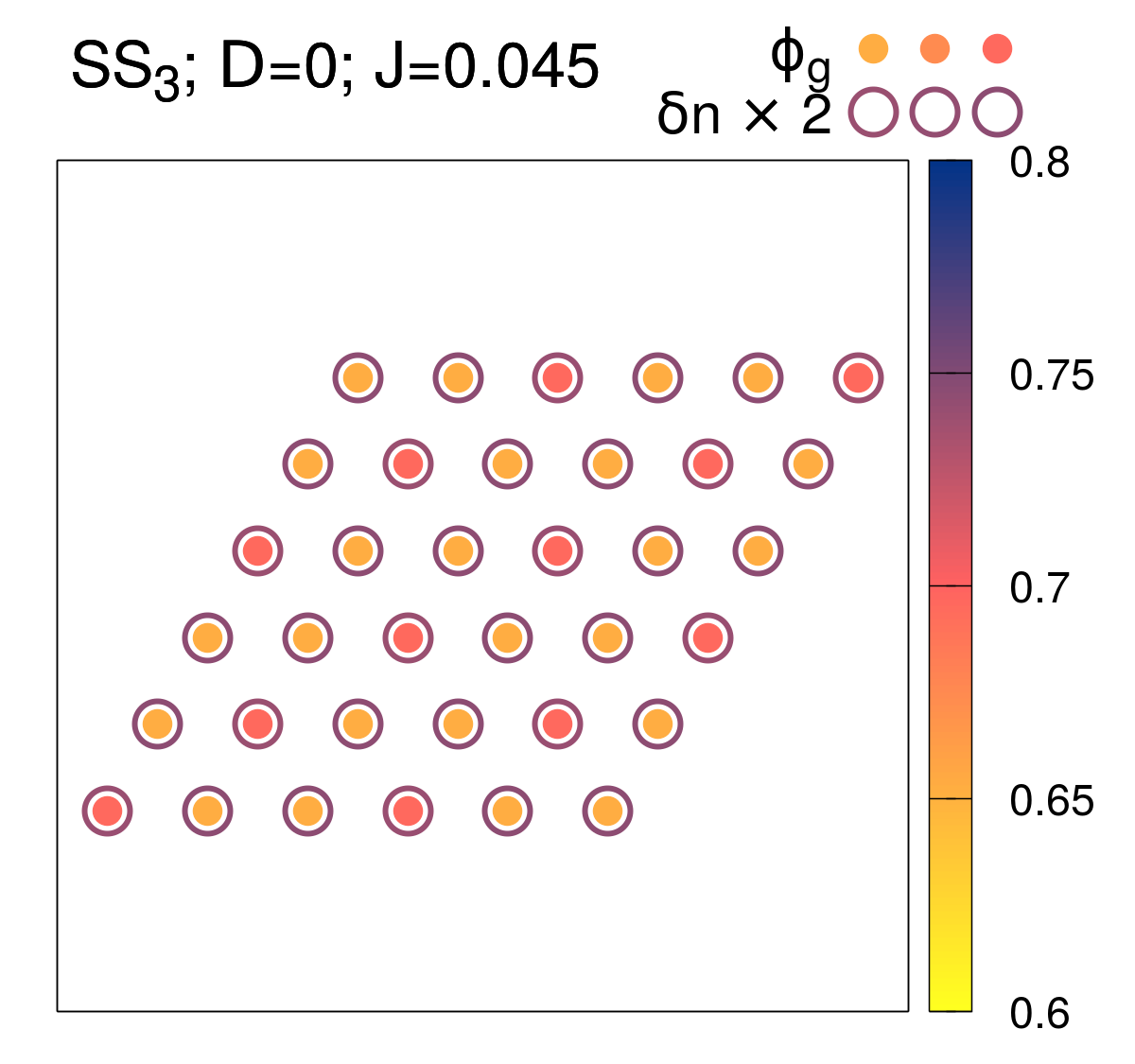}
\includegraphics[width=0.32\linewidth]{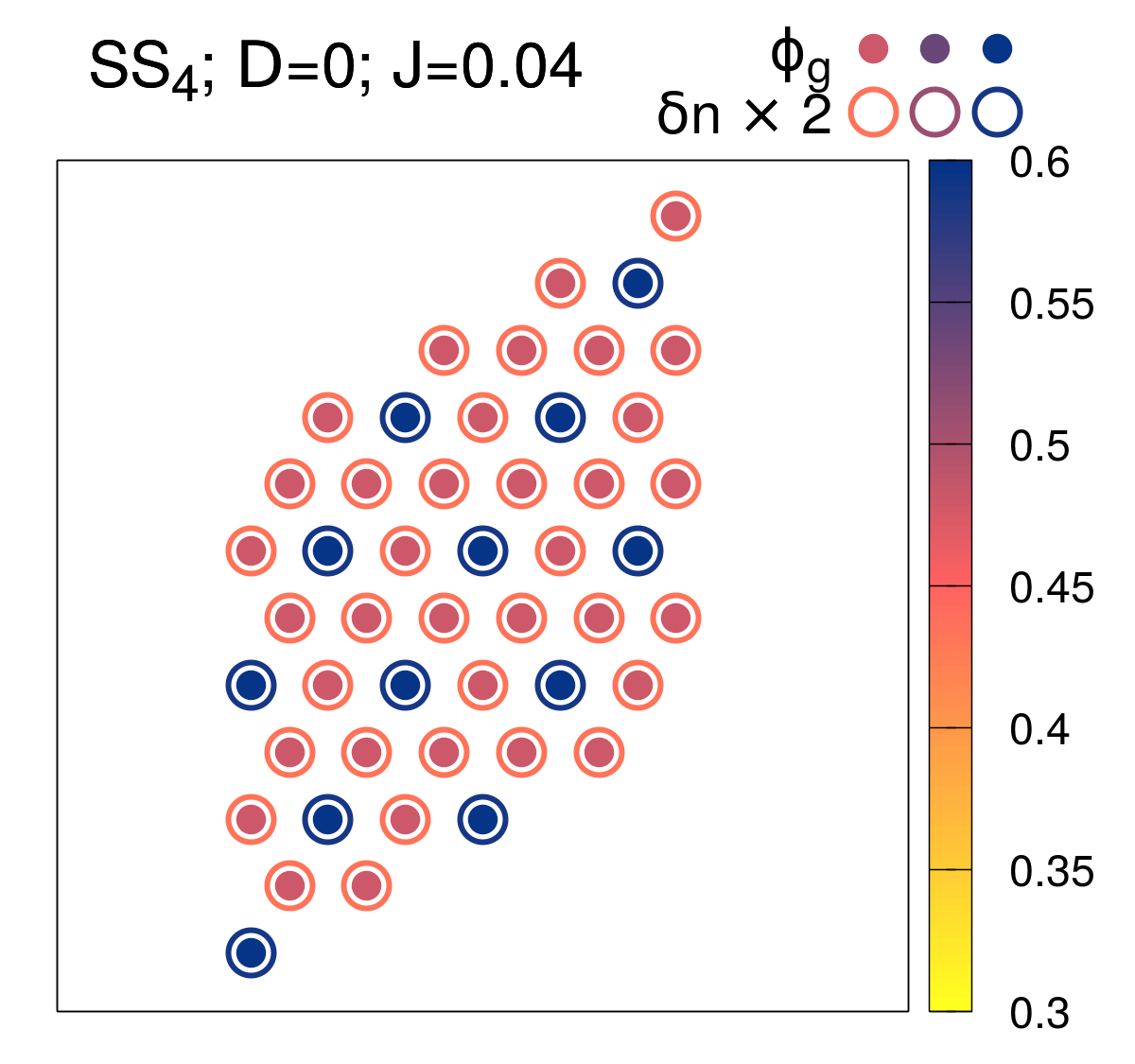}
\includegraphics[width=0.32\linewidth]{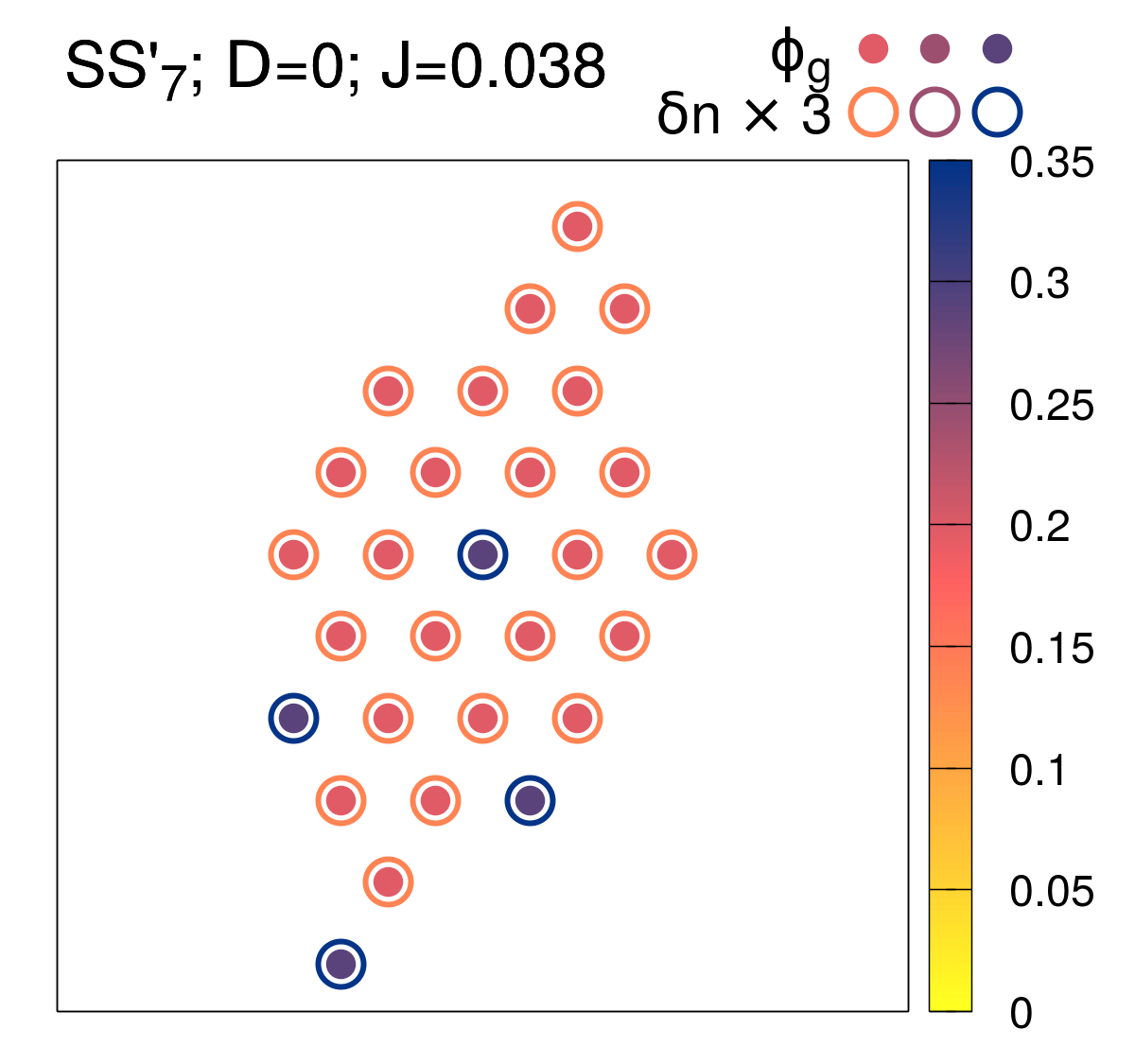}
\caption{\label{fig:struct2} Local condensate order parameter of ground state bosons $\phi_{i,g}=\langle \hat{a}_i\rangle$ (filled circles) and local number fluctuations $\delta n^2_i = \langle \hat{n}_i^2 \rangle - \langle \hat{n}_i \rangle^2$ (empty circles) for different phases observed in Fig.~\ref{fig:phdiag_BDMFT}. Here $\hat{n}_i = \hat{n}_{g,i}+\hat{n}_{e,i}$. Each graph represents sites within a quadrupled unit cell in the same way as in Fig.~\ref{fig:struct}. Note the ranges used for the $\mathrm{SS}_3$ and $\mathrm{SS}_4$ phases, and the rescaling of the magnitude of fluctuations used for better visibility.}
\end{figure*}

\section{Supplementary results}
\label{app:supplementary results}

\subsection{Self-consistency test}

In order to determine the accuracy of our assumption regarding the self-consistency condition~\eqref{eq:betheSC} described in Sec.~\ref{sec:sub2sec2} we have implemented the full self-consistency condition and performed calculations for limited range of parameters. We have chosen to set the detuning to $\Delta=0$ and vary the hopping amplitude $J$ as this gives a cross section of the most interesting part of the phase diagram shown in Fig.~\ref{fig:phdiag_BDMFT}. We compare the average occupation of the excited state bosons as a function of $J$. The result is shown in Fig.~\ref{fig:SC_test}.

One can see that for the majority of values of the hopping amplitude $J$ the two self-consistency equations yield quantitatively comparable results. The major difference appears in the region of $J\in[0.031,0.03925]$.  In the simplified self-consistency the extent of $\mathrm{DW}'_7$ phase is much smaller while the extent of $\mathrm{SS}'_7$ phase is slightly larger. In the latter phase the simplified approach yields also larger rate of change of the average occupation of excited state bosons with increasing $J$. We note that while the extent of these phases is affected, the general features of the phase diagram remain unaffected. We do not observe significant differences, e.g. appearance of new types of phases. Tests for other values of $\Delta$, not shown here, confirmed this conclusion.

\subsection{Condensate order parameter and local fluctuations}

In Fig.~\ref{fig:struct2} we present additional results showing local condensate fraction $\phi_{i,g}=\langle \hat{a}_i\rangle$ and local fluctuations of site occupation $\delta n^2_i=\langle \hat{n}_i^2 \rangle - \langle \hat{n}_i \rangle^2$, for different sites $i$. In the phases $\mathrm{SS}_7$, $\mathrm{SS}_9$ and $\mathrm{SS}_{12}$ we observe that both the condensate order parameter and local fluctuations are significantly suppressed at the sites occupied by the excited state bosons. This is consistent with our interpretation that in these phases we observe frozen bosons on selected sites with non-vanishing excited state fraction and condensed bosons in the intermediate sites, which are responsible for the superflow. A qualitatively different behavior is observed for the $\mathrm{SS}_3$, $\mathrm{SS}_4$ and $\mathrm{SS}'_7$ phases. There the condensate fraction and local fluctuations are actually larger at the sites with a significant excited state bosons fraction. This property is reminiscent of the bubble supersolids observed in Ref.~\cite{henkel2012,cinti2014}, although here we work in the significantly different regime of small detuning. We note that the spatial modulation of the two observables considered here is much smaller in the $\mathrm{SS}_3$ and $\mathrm{SS}_4$ phase than in the $\mathrm{SS}'_7$ phase.

\bibliography{frustrated_lattice}

\end{document}